\documentclass[12pt, draftclsnofoot, onecolumn]{IEEEtran}
\usepackage[T1]{fontenc}
\usepackage[cmintegrals]{newtxmath}

  \usepackage{amsmath,amssymb}
  
  \usepackage{bbm,bm}
  \usepackage{pgfplots}
  \usepackage{siunitx}
  \usepackage{graphicx}
  \usepackage{algorithm,algpseudocode}
  \usepackage{mathtools}
  \usepackage{url}
  \usepackage[export]{adjustbox}
\usepackage{color}
\usepackage{todonotes}
\usepackage{setspace} 
\ifCLASSOPTIONcompsoc
 \usepackage[caption=false,font=normalsize,labelfont=sf,textfont=sf]{subfig}
\else
 \usepackage[caption=false,font=footnotesize]{subfig}
\fi

\usepackage{tikz,tkz-graph}
\usepackage[mode=buildnew]{standalone}
\usetikzlibrary{matrix,positioning,fit,calc,shapes}
\pgfplotsset{compat=newest}
\algnewcommand{\Initialize}[1]{%
  \State \textbf{Initialize}
  \Statex \hspace*{\algorithmicindent}\parbox[t]{0.8\linewidth}{\raggedright #1}
}

\let\oldbrace\{
\def\{{\oldbrace\kern0.5pt}

\newcommand{\Rc}{\mathcal{R}}

\newcommand{\Xc}{\mathcal{X}}
\newcommand{\Yc}{\mathcal{Y}}

\newcommand{\yv}{\boldsymbol{y}}

\newcommand{\sv}{\boldsymbol{s}}

\newcommand{\LLR}{\mathrm{LLR}}

\newcommand{\tikzmark}[2]{\tikz[overlay,remember picture,
  baseline=(#1.base)] \node (#1) {#2};}
\newcommand{\argmax}{\operatornamewithlimits{argmax}}
\newcommand{\ve}{\bm}
\newtheorem{theorem}{Theorem}
\newtheorem{proposition}{Proposition}
\newtheorem{remark}{Remark}
\newtheorem{lemma}{Lemma}
\newtheorem{example}{Example}

\definecolor{ins}{cmyk}{0.5,0.3,0,0.1}

\begin{document}
\allowdisplaybreaks

\title{Compute--Forward Multiple Access (CFMA):\\Practical Code Design}

%

\author{Erixhen~Sula,~\IEEEmembership{Student Member,~IEEE,}
	     Jingge~Zhu,~\IEEEmembership{Member,~IEEE,}
	     Adriano~Pastore,~\IEEEmembership{Member,~IEEE,}
	     Sung~Hoon~Lim,~\IEEEmembership{Member,~IEEE,}
	     and~Michael~Gastpar,~\IEEEmembership{Fellow,~IEEE}
\thanks{E. Sula and M. Gastpar are with the School of Computer and Communication Sciences, EPFL, Lausanne, 1015, Switzerland (e-mail: erixhen.sula@epfl.ch; michael.gastpar@epfl.ch).}
\thanks{J. Zhu is with the department of electrical engineering and computer sciences, UC Berkeley, 94720 Berkeley, USA (e-mail: jingge.zhu@berkeley.edu)}
\thanks{A. Pastore is with the CTTC, 08860 Castelldefels, Spain (e-mail: adriano.pastore@cttc.cat)}
\thanks{S. H. Lim is with the KIOST, 49111 Busan, Korea (e-mail: shlim@kiost.ac.kr)}
}

\maketitle

\begin{abstract}
We present a practical strategy that aims to attain rate points on the dominant face of the multiple access channel capacity using a standard low complexity decoder. 
This technique is built upon recent theoretical developments of Zhu and Gastpar on compute-forward multiple access (CFMA) which achieves the capacity of the multiple access channel using a sequential decoder. We illustrate this strategy with off-the-shelf LDPC codes. In the first stage of decoding, the receiver first recovers a linear combination of the transmitted codewords using the sum-product algorithm (SPA). In the second stage, by using the recovered sum-of-codewords as side information, the receiver recovers one of the two codewords using a modified SPA, ultimately recovering both codewords. The main benefit of recovering the sum-of-codewords instead of the codeword itself is that it allows to attain points on the dominant face of the multiple access channel capacity without the need of rate-splitting or time sharing while maintaining a low complexity in the order of a standard point-to-point decoder. This property is also shown to be  crucial for some applications, e.g., interference channels. For all the simulations with single-layer binary codes, our proposed practical strategy is shown to be within \SI{1.7}{\decibel} of the theoretical limits, without explicit optimization on the off-the-self LDPC codes.
\end{abstract}


\section{Introduction}
\label{sec:intro}


\IEEEPARstart{T}{he} interference channel \cite{Han--Kobayashi1981} has been considered as a canonical model to understand the design principles of cellular networks with inter-cell interference. For example, the Han-Kobayashi scheme~\cite{Han--Kobayashi1981,Etkin--Tse--Wang2008} has been developed for the interference channel and is currently the best known strategy for the two-user interference channel. The Han-Kobayashi scheme operates by superposition encoding and sending two code components, a private part and a common part. Both components are optimally recovered by using simultaneous joint decoding~\cite{Chong--Motani--Garg--El-Gamal2008, Bandemer--El-Gamal--Kim2012}. While the theoretical aspects of this strategy is well understood, how to practically implement it with low complexity remains as an important question. Since recovering the private part can be understood as a simple application of point-to-point codes, 
the main challenge is to send the common messages, i.e., each receiver treats the interference channel as a {\em multiple access channel} and recovers {\em both} messages (decoding interference). The goal of this work is to design practical low-complexity decoders that can closely attain the promised performance of simultaneous decoding over multiple access channels.

Compared to the multiple access channel, the major difficulty in designing practical strategies for decoding interference in an interference channel is the following. In a multiple access channel, it is well known that sequential decoding (e.g. successive cancellation) and time sharing or rate-splitting can achieve the capacity region of the multiple--access channel~\cite{Rimoldi--Urbanke1996, Grant--Rimoldi--Urbanke--Whiting2001,El-Gamal--Kim2011}. However, this is not the case when we treat the interference channel as a combination of {\em two} MACs (compound MAC), that is, a channel in which the pair of encoders send messages to both decoders. It was well observed in~\cite{Fawzi--Savov2012, Wang--Sasoglu--Kim2014} that successive cancellation decoding with time sharing is strictly sub-optimal compared to joint decoding for this case.

Recently, there has been some advances in designing low-complexity sequential decoding strategies that can overcome this issue.
In particular, Wang, \c{S}a\c{s}o\u{g}lu, and Kim~\cite{Wang--Sasoglu--Kim2014} presented the sliding-window superposition coding (SWSC) strategy that achieves the performance of simultaneous decoding with existing point-to-point codes. A case study using standard LTE codes of the SWSC scheme was given in~\cite{Park--Kim--Wang2014,Kim--Ahn--Kim--Park--Wang--Chen--Park2015}. An important component of this strategy is that they use block Markov encoding which requires multi-block transmissions. In another line of work, Zhu and Gastpar~\cite{Zhu--Gastpar2014b} proposed the compute--forward multiple access (CFMA) strategy based on the compute--and--forward framework proposed in \cite{Nazer--Gastpar2011}. It is shown that CFMA achieves the capacity region of the Gaussian MAC with sequential decoding when the SNR of both decoders is greater than $1+\sqrt{2}$, and the strategy is also extended to the interference channel in \cite{zhu_lattice_2015}. Under this condition, the optimal performance is achieved using single block transmissions. The main component of CFMA is that it utilizes lattice codes to first compute a linear combination of the codewords sent by the transmitters, which is accomplished by extending the compute--and--forward strategy originally proposed by Nazer and Gastpar~\cite{Nazer--Gastpar2011}. In the next stage, by using the linear combination of the codewords as side information, the decoder recovers any one of the messages\footnote{In general the CFMA strategy finds and computes the best two linearly independent combinations.}. By having one linear combination and one of the other messages, the receiver can recover both messages. By appropriately scaling the lattice codes, it is shown in~\cite{Zhu--Gastpar2014b} that the dominant face of the capacity region of the multiple access channel can be attained with sequential decoders without time sharing, and thus, is optimal for the compound MAC setting under the SNR condition. Following the theoretical study of CFMA~\cite{Zhu--Gastpar2014b} which was attained with  lattice codes and lattice decoding~\cite{Erez--Zamir2004}, the main goal of our work is to design practical codes and efficient decoding algorithms that can attain the achievable rates of CFMA. To accomplish this goal and for applications to current systems, we will use off-the-shelf {\em low-density parity check} (LDPC) codes for point-to-point communications as our basic code component. However, we emphasize the fact that this technique can be applied in conjunction with any linear codes, given appropriate modifications. Several works on the design of practical compute--forward strategies have also been considered in~\cite{Ordentlich--Zhan--Erez--Gastpar--Nazer2011, Hern--Narayanan2013,Feng--Silva--Kschischang2013,Mejri--Othman2015}. The main difference from the previous works is that we design a practical compute--forward strategy by explicitly using linear codes and the sum-product decoding algorithm. 

Our main contribution in this work is summarized as follows:

\begin{itemize}
\item Following the theoretical insights developed in~\cite{Lim--Feng--Pastore--Nazer--Gastpar2016} for compute--and--forward using nested linear codes, we design {\em nested} LDPC codes based on off-the-shelf LDPC codes for a practical implementation of CFMA.
\item Our decoding algorithm which is designed in a successive manner has complexity in the order of a conventional sum-product decoder. In the first step of the decoding algorithm, we recover a linear combination of LDPC codewords using the sum-product algorithm (SPA). In the next stage, we recover a codeword with another sum-product decoder using the previously recovered linear combination as side information. In order to adapt the standard SPA in our scenario, we need to modify the initial input log-likelihood ratio (LLR) accordingly for each decoding step. 
\item To support higher data rates using existing off-the-shelf binary codes, we propose a {\em multilevel binary} LDPC code construction with higher order modulation. By appropriately modifying the input log likelihood ratio (LLR), we show that the standard SPA can be used in the decoding procedure. In particular, the standard SPA will be applied to every bit in the binary representation of the codeword separately with a modified initial input LLR, and the previously decoded bits will be used as side information for decoding the bits in the next level. 
\item We further extend our technique to more complicated scenarios. Specifically,  we discuss how to adapt our implementation to complex-valued channel models, to multiple-access channels with more than two users, and to interference channels.

\end{itemize}

The remainder of the paper is organized as follows. In the next section we present our system model and discuss the theoretical background motivating our work. In section \ref{sec:CFMA basic} the basic CFMA with binary codes is studied. In section \ref{sec:CFMA extensions} extension of CFMA with multilevel binary codes is examined in details. In section \ref{sec:CFMA complex} extension to complex channel is treated. Section \ref{sec:simulation} provides simulations results for various scenarios.

We use boldface lower-case and upper-case to denote vectors and matrices respectively. The operator `$\oplus_q$' denotes $q$-ary addition, and `$\oplus$' (without subscript) is to be understood as binary addition ($q=2$). The bracket $[a:b]$ denotes the set of integers $\{a,a+1,\ldots, b\}$.

\section{System Model and Theoretical Background} \label{sec:system model}
We consider the two-user Gaussian MAC, with input alphabets $\Xc_1$, $\Xc_2$ and output alphabet $\Yc$. For input symbols $x_1$, $x_2$, the output symbol is given by
	$Y
	 = h_1 x_1 + h_2 x_2 + Z,$
where $h_1,h_2 \in \mathbb{R}$ denote the constant channel and $
Z\sim\mathcal{N}(0,1)$ is the additive white Gaussian noise.

\begin{figure}[ht]	
\centering
\begin{tikzpicture}[every path/.append style={thick}]
\matrix[matrix of math nodes, row sep=1cm, inner sep=2mm, nodes=draw] (enc) {
	\mathcal{E}_1 \\
	\mathcal{E}_2 \\
};
\node[draw,shape=circle,inner sep=.5mm,right=12mm of enc] (adder) {$+$};
\draw[<-] (enc-1-1.west)--+(-5mm,0) node[left] {$w_1$};
\draw[<-] (enc-2-1.west)--+(-5mm,0) node[left] {$w_2$};
\draw[->] (enc-1-1.east)--+(7mm,0) node[midway,above] {$\bm{x}_1$} -- (adder) node[midway,above] {$h_1$};
\draw[->] (enc-2-1.east)--+(7mm,0) node[midway,above] {$\bm{x}_2$} -- (adder) node[midway,below] {$h_2$};
\draw[<-] (adder.south)--+(0,-5mm) node[below] {$\bm{z}$};
\node[draw, right=1cm of adder, inner sep=2mm] (dec) {$\mathcal{D}$};
\draw[->] (adder.east)--(dec.west) node[midway,above] {$\bm{y}$};
\draw[->] (dec.east)--+(5mm,0) node[right] {$(\hat{w}_1,\hat{w}_2)$};
\end{tikzpicture}
\caption{Block diagram of the two-user Gaussian MAC communication system}
\label{fig:graph_repre}
\end{figure}
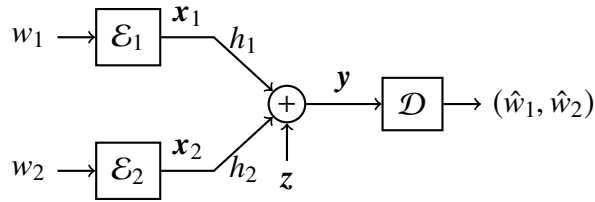

A $(2^{n R_1},2^{n R_2},n)$ code for the MAC consists of:

\begin{itemize}
\item two message sets $\bigl[1:2^{n R_1}\bigr]$ and $\bigl[1:2^{n R_2}\bigr]$;
                          \item two encoding functions $\mathcal{E}_1$ and $\mathcal{E}_2$ which assign codewords $\ve{x}_k\in\Xc_k^n$ to each message $w_k \in \bigl[1:2^{n R_k}\bigr]$, $k=1,2$, with average power constraint 
$\|\ve{x}_k\|^2 \le n P$;
\item a decoding function $\mathcal{D}$ which assigns an estimate $(\hat{w}_1,\hat{w}_2)$ of the message pair based on $\ve{y}\in\Yc^n$.
\end{itemize}
Assuming that the messages $w_1$ and $w_2$ are drawn uniformly at random from the message sets $\bigl[1:2^{n R_1}\bigr]$ and $\bigl[1:2^{n R_2}\bigr]$, respectively, the average probability of error is defined as
\begin{equation}
	P_\mathrm{e}^{(n)}
	 = \mathrm{Prob}\bigl\{(\hat{w}_1,\hat{w}_2) \neq (w_1,w_2)\bigr\}.
\end{equation}

A rate pair $(R_1,R_2)$ is said to be \emph{achievable} if there exists a sequence of $(2^{n R_1},2^{n R_2},n)$ codes such that $\lim_{n \rightarrow \infty} P_\mathrm{e}^{(n)}=0$.

In the remainder of this section, we will briefly review some theoretical results on the multiple--access channel and the compute--forward multiple access (CFMA) scheme.

\begin{figure}[ht]
\centering
\begin{tikzpicture}[every path/.append style={thick}]
\begin{axis}[
	xmin=0, xmax=1.4,
	ymin=0, ymax=1.2,
	xticklabels={}, yticklabels={}, ticks=none,
	xlabel={$R_1$}, ylabel={$R_2$},
	axis lines=middle,
	width=7cm,
	axis equal image,
	axis on top=true,
	clip=false,
	]
    \coordinate (origin) at (0,0);
	\coordinate (A) at (.6,1);
	\coordinate (B) at (1.2,.4);
\fill[fill=black!10] (origin)--(B |- origin)--(B)--(A)--(A -| origin)--cycle;
\fill[fill=black!20] (origin)--(B |- origin)--(B)--(A |-B)--(A)--(A -| origin)--cycle;
\draw[dashed] (A -| origin)--(A)--(A |- origin);
\draw[dashed] (B -| origin)--(B)--(B |- origin);
\node[draw, fill=white, shape=circle, inner sep=1.2pt, label={[label distance={-.5ex}]60:A}] at (A) {};
\node[draw, fill=white, shape=circle, inner sep=1.2pt, label={[label distance={-.5ex}]30:B}] at (B) {};
\draw[<-] ($(A)!.5!(B)$) --+ (.08,.08) 
node[align=left,anchor=south west] {\small dominant\\face};
\end{axis}
\end{tikzpicture}
\caption{The MAC rate region achievable for a fixed input distribution $p(x_1)p(x_2)$ (light and dark shaded gray combined). The corner points $\mathsf{A}$ and $\mathsf{B}$ are achievable by successive cancellation and points located on the dominant face (i.e., the segment connecting $\mathsf{A}$ and $\mathsf{B}$) can be achieved by time sharing between $\mathsf{A}$ and $\mathsf{B}$.}
\label{fig:mac-cap}
\end{figure}
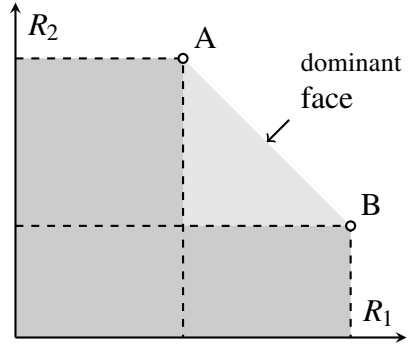

The capacity region of the MAC~\cite{Cover--Thomas2006}
is given by the convex hull of the set of rate pairs $(R_1, R_2)$ that satisfies
\begin{subequations}\label{eq:mac-cap}
\begin{align}
R_1&<I(X_1; Y|X_2)\\
R_2&<I(X_2; Y|X_1)\\
R_1+R_2&<I(X_1, X_2; Y)
\end{align}
\end{subequations}
for some joint distribution $(X_1,X_2) \sim p(x_1)p(x_2)$. When input alphabets are set to $\mathcal{X}_1 = \mathcal{X}_2 = \mathbb{R}$ and the codewords are subject to a power constraint $P$, the capacity region~\cite{Cover--Thomas2006} is attained by setting $X_k\sim\mathcal{N}(0,P)$, $k=1,2$. 
\begin{remark}
In this work, our practical code design for the Gaussian multiple access channel will be based on uniform discrete input distributions. As a target reference to this case, we will often compare our practical results with the achievable rate region of \eqref{eq:mac-cap} evaluated with uniformly distributed discrete inputs. In this case, we will refer to the rate region~\eqref{eq:mac-cap} with uniform discrete input distributions as $\Rc_{\sf MAC-UI}$.
\end{remark}

The so-called {\em corner points}, labeled in Figure~\ref{fig:mac-cap} as $\mathsf{A}$ and $\mathsf{B}$, are achievable by means of successive cancellation decoding~\cite{Cover--Thomas2006}.
The remaining points on the dominant face (that is, the segment connecting corner points) can be achievable by time-sharing.

However, in some networks such as interference channels, successive cancellation with time sharing can result in suboptimal performance such as the interference channel example discussed in Section \ref{sec:intro}. For the strong interference case, the best approach is to recover both messages at the decoders, i.e., we treat the interference channel as a compound MAC. The main challenge in designing low complexity strategies for the compound MAC thus lies in the ability to attain the rate points on the dominant face of the multiple access rate region \emph{without time sharing or rate-splitting}.

Recently, Zhu and Gastpar \cite{Zhu--Gastpar2014b} proposed a novel low-complexity encoding and decoding scheme for the Gaussian MAC that does not require time-sharing or rate-splitting, yet achieves all points on the dominant face under the mild condition 
$\frac{h_1h_2P}{\sqrt{1+h_1^2P+h_2^2P}}\geq 1.$
This compute-and-forward \cite{Nazer--Gastpar2011} based scheme employs nested lattice codes and a sequential compute-and-forward decoding scheme: in the first step, the receiver decodes a linear codeword combination $a_1 \ve{u}_1 + a_2 \ve{u}_2$ (modulo-lattice reduced) with non-zero integer coefficients $a_1$ and $a_2$; in the second step, the receiver exploits this linear combination as side information to decode one of the codewords (either $\ve{u}_1$ or $\ve{u}_2$); finally, the receiver reconstructs the other message from additively combining the outputs of the previous two decoding steps. Furthermore, it is shown in \cite{zhu_lattice_2015} that for the interference channel with strong interference (or the compound MAC), this decoding scheme achieves (a part of) the  dominant face of the capacity region.

More recently, the authors of~\cite{Lim--Feng--Pastore--Nazer--Gastpar2016} presented a generalization of compute--forward which is based on nested linear codes and joint typicality encoding and decoding, rather than on lattice codes as in \cite{Nazer--Gastpar2011,Zhu--Gastpar2014b}. In this setup, one defines field mappings $\varphi^{-1}_1$ and $\varphi^{-1}_2$ which bijectively map constellation points $x_1 \in \mathcal{X}_1$ and $x_2 \in \mathcal{X}_2$ to finite field elements $u_1 \in \mathbb{F}_q$ and $u_2 \in \mathbb{F}_q$, respectively. In analogy to the procedure with lattice codes described above, the receiver first decodes a weighted sum
	$\ve{s}
    = a_1 \ve{u}_1 \oplus_q a_2 \ve{u}_2 = a_1 \varphi^{-1}_1(\ve{x}_1) \oplus_q a_2 \varphi^{-1}_2(\ve{x}_2)$
(where $\varphi^{-1}_k(\ve{x}_k)$ stands for the vector resulting from a symbol-by-symbol application of $\varphi^{-1}_k$) and then uses the latter as side information to decode either $\ve{u}_1$ or $\ve{u}_2$.

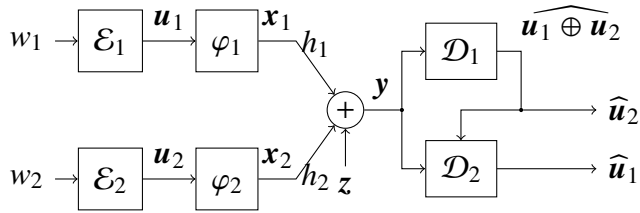
\begin{figure}[ht]
\centering
\begin{tikzpicture}
\matrix[matrix of math nodes, row sep=1cm, column sep=7mm, inner sep=2mm, nodes={draw,anchor=base, text height=.8em, text depth=.2em}] (modenc) {
	\mathcal{E}_1 & \varphi_1 \\
	\mathcal{E}_2 & \varphi_2 \\
};
\node[draw,shape=circle,inner sep=.5mm,right=7mm of modenc] (adder) {$+$};
\draw[<-] (modenc-1-1.west)--+(-3mm,0) node[left] {$w_1$};
\draw[<-] (modenc-2-1.west)--+(-3mm,0) node[left] {$w_2$};
\draw[->] (modenc-1-1.east)--(modenc-1-2.west) node[midway,above] {$\ve{u}_1$};
\draw[->] (modenc-2-1.east)--(modenc-2-2.west) node[midway,above] {$\ve{u}_2$};
\draw[->] (modenc-1-2.east)--+(5mm,0) node[midway,above] {$\ve{x}_1$} -- (adder) node[midway,above] {$h_1$};
\draw[->] (modenc-2-2.east)--+(5mm,0) node[midway,above] {$\ve{x}_2$} -- (adder) node[midway,below] {$h_2$};
\draw[<-] (adder.south)--+(0,-5mm) node[below] {$\ve{z}$};
\matrix[matrix of math nodes, row sep=8mm, column sep=7mm, inner sep=2mm, nodes=draw, right=6mm of adder] (dec) {
	\mathcal{D}_1 \\
	\mathcal{D}_2 \\
};
\node[circle,draw=none,fill,inner sep=0,outer sep=0,minimum height=4\pgflinewidth] (branch) at ($(adder.east)+(5mm,0)$) {};
\node[circle,draw=none,fill,inner sep=0,outer sep=0,minimum height=4\pgflinewidth] (branch2) at ($(dec-1-1.south)!.5!(dec-2-1.north) + (8mm,0)$) {};
\draw (adder.east)--(branch) node[midway,above] {$\ve{y}$};
\draw[->] (branch)|-(dec-1-1.west);
\draw[->] (branch)|-(dec-2-1.west);
\draw[->] (dec-1-1.east) -| (branch2) node[near start,above right] {$\widehat{\ve{u}_1 \oplus \ve{u}_2}$} -| (dec-2-1.north);
\coordinate (tip) at ($(branch2)+(10mm,0)$);
\draw[->] (dec-2-1.east)--(dec-2-1.east -| tip) node[right] {$\widehat{\ve{u}}_1$};
\draw[->] (branch2) -- (tip) node[right] {$\widehat{\ve{u}}_2$};
\end{tikzpicture}
\caption{Block diagram for the MAC channel with CFMA decoding.}
\end{figure}

By specializing~\cite[Theorem~1]{Lim--Feng--Pastore--Nazer--Gastpar2016}, we can readily establish the following theorem which describes a rate region achievable with CFMA.

Let $\mathcal{X}_1$ and $\mathcal{X}_2$ have equal cardinality $q$, which we assume to be a prime power. Let $\mathbb{F}_q$ denote a $q$-ary finite field. Fix the input distribution $p(x_1)p(x_2)$. 
\begin{theorem}[\cite{Lim--Feng--Pastore--Nazer--Gastpar2016}]   \label{thm:cfma-nlc}
A rate pair $(R_1, R_2)$ is achievable with nested linear codes and under the CFMA decoding strategy\footnote{The reader is referred to \cite{Lim--Feng--Pastore--Nazer--Gastpar2016} for a precise description of the nested linear code construction, and the encoding and decoding strategies used for the proof of achievability.} if for some non-zero coefficient vector $(a_1,a_2) \in \mathbb{F}_q^2$ and for some bijective field mappings $\varphi^{-1}_1$ and $\varphi^{-1}_2$, we have either
\begin{subequations} \label{eq:cfma-nlc}
\begin{equation}   \label{A_prime}
\begin{split}
	R_1 &\leq  H(X_1) - \max\bigl\{ H(S|Y),\,H(X_1,X_2|Y,S) \bigr\}, \\
	R_2 &\leq H(X_2) - H(S|Y),
\end{split}
\end{equation}
or
\begin{equation}   \label{B_prime}
\begin{split}
	R_1 &\leq H(X_1) - H(S|Y), \\
	R_2 &\leq H(X_2) - \max\bigl\{ H(S|Y),\,H(X_1,X_2|Y,S) \bigr\},
\end{split}
\end{equation}
\end{subequations}
where $S = a_1 U_1 \oplus_q a_2 U_2$ and $(U_1,U_2) = (\varphi^{-1}_1(X_1),\varphi^{-1}_2(X_2))$.
\end{theorem}
Throughout the paper, we explicitly denote the rate region (\ref{eq:cfma-nlc}) evaluated with uniform discrete inputs by $\Rc_{\sf CFMA-UI}$.

\begin{figure}[ht]
\centering
\begin{tikzpicture}[every path/.append style={thick}]
\begin{axis}[
	xmin=0, xmax=1.4,
	ymin=0, ymax=1.2,
	xticklabels={}, yticklabels={}, ticks=none,
	xlabel={$R_1$}, ylabel={$R_2$},
	axis lines=middle,
	width=7cm,
	axis equal image,
	axis on top=true,
	clip=false,
	]
    \coordinate (origin) at (0,0);
	\coordinate (A) at (.6,1);
	\coordinate (B) at (1.2,.4);
	\coordinate (AA) at (.7,.9);
	\coordinate (BB) at (.9,.7);
	\fill[black!10] (origin)-|(B)--(A)-|(origin);
	\fill[black!20] (origin)-|(BB)-|(AA)-|(origin);
	\draw[dashed]
    	(AA -| origin)--(AA)--(AA |- origin)
    	(BB -| origin)--(BB)--(BB |- origin);
	\node[draw,fill=white,shape=circle,inner sep=1.2pt,label={[label distance={-.5ex}]60:\textsf{A}}] at (A) {};
	\node[draw,fill=white,shape=circle,inner sep=1.2pt,label={[label distance={-.5ex}]60:\textsf{A'}}] at (AA) {};
	\node[draw,fill=white,shape=circle,inner sep=1.2pt,label={[label distance={-.5ex}]30:\textsf{B}}] at (B) {};
	\node[draw,fill=white,shape=circle,inner sep=1.2pt,label={[label distance={-.5ex}]30:\textsf{B'}}] at (BB) {};
    \draw[dotted] (origin)--(1.1,1.1);
	\draw (.2,0) arc (0:45:.2) node[midway, right] {\ang{45}};
\end{axis}
\end{tikzpicture}
\caption{For some fixed $(a_1,a_2) \in \mathbb{F}_q^2$, inequalities \eqref{A_prime} and \eqref{B_prime} yield rate points $\mathsf{A}'$ and $\mathsf{B}'$, respectively. If $H(S|Y) \leq \tfrac{1}{2} H(X_1,X_2|Y)$, they lie on the dominant face (as in the Figure). Points dominated by $\mathsf{A}'$ and $\mathsf{B}'$ (in dark gray) can be achieved without time sharing, using nested linear codes and the CFMA decoding strategy. The uniform-input rate region $\Rc_{\sf MAC-UI}$ is shaded in light gray.}
\label{fig:mac-nlc}
\end{figure}
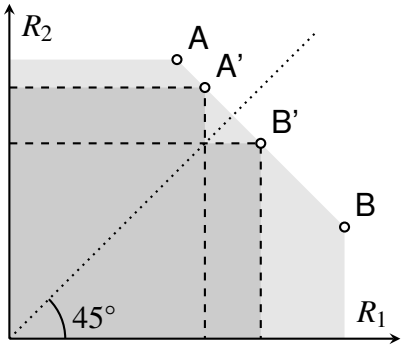

Note that Theorem~\ref{thm:cfma-nlc} can be extended by a discretization approach \cite[Theorem~3]{Lim--Feng--Pastore--Nazer--Gastpar2016} to infinite constellations and continuous signal distributions, by way of which, in particular, Zhu and Gastpar's original achievability result proved using lattice codes \cite[Theorem~2]{Zhu--Gastpar2014b} can be recovered.

Figure~\ref{fig:mac-nlc} illustrates the rates achievable with CFMA according to Theorem~\ref{thm:cfma-nlc}, for some fixed coefficient pair $(a_1,a_2) \in \mathbb{F}_q^2$.
The coordinates of points $\mathsf{A}'$ and $\mathsf{B}'$ are given by the right-hand sides of Equations~\eqref{A_prime} and \eqref{B_prime}, respectively.
One can show that $\mathsf{A}'$ is located on the dominant face and is distinct from $\mathsf{A}$ if and only if
\begin{subequations}
\begin{equation}   \label{A_prime_on_DF}
	H(X_2|Y) < H(S|Y) \leq \tfrac{1}{2}H(X_1,X_2|Y).
\end{equation}
Similarly, $\mathsf{B}'$ is located on the dominant face and is distinct\footnote{If we allow $\mathsf{A}'$ and $\mathsf{B}'$ to coincide with corner points, strict inequalities in \eqref{A_prime_on_DF} and \eqref{B_prime_on_DF} have to be replaced by weak inequalities.} from $\mathsf{B}$ if and only if
\begin{equation}   \label{B_prime_on_DF}
	H(X_1|Y) < H(S|Y) \leq \tfrac{1}{2}H(X_1,X_2|Y).
\end{equation}
\end{subequations}
Additionally, if---as we shall assume throughout the article---the auxiliaries $U_1$ and $U_2$ are i.i.d.\ uniform on $\mathbb{F}_q$ (or equivalently, if $X_1$ and $X_2$ are uniform over their respective constellations), then $\mathsf{A}'$ and $\mathsf{B}'$ are reflections of one another about the symmetric rate line (dotted line), like in Figure~\ref{fig:mac-nlc}.
\begin{example}
For the binary field $\mathbb{F}_2 = \{0,1\}$, the coefficient pair can be either $(1,1)$, $(0,1)$ or $(1,0)$. Provided that $U_1$ and $U_2$ are i.i.d.\ uniform, by evaluation of \eqref{A_prime}--\eqref{B_prime}, the pairs $(0,1)$ and $(1,0)$ each yield one of the two corner points $\mathsf{A}$ or $\mathsf{B}$, and at those corner points, CFMA decoding reduces to conventional successive decoding. By contrast, the coefficients $(a_1,a_2)=(1,1)$ yield a pair of rate points $\mathsf{A}'$ and $\mathsf{B}'$ that lie on the dominant face and are distinct from corner points, much like the situation depicted in Figure~\ref{fig:mac-nlc}.
\end{example}

For field sizes $q>2$, as the coefficient pair $(a_1,a_2)$ is varied over $\mathbb{F}_q^2$, more pairs of points $\mathsf{A}'$ and $\mathsf{B}'$ located on the dominant face may be attained. In the limiting case of $q \to \infty$, a continuous subset of the dominant face may be achieved with CFMA, as exemplified by \cite[Theorem~3]{Zhu--Gastpar2014b}.

In the following sections, we will propose practical CFMA strategies using nested LDPC codes inspired by the theoretically achievable rate region proposed by Theorem~\ref{thm:cfma-nlc}.


\section{CFMA with binary codes}
\label{sec:CFMA basic}

In this section we devise a practical design of CFMA for a two-user Gaussian MAC, based on off-the-shelf binary linear error-correcting codes. One important feature of the proposed implementation is that while operating near the dominant face of the achievable rate region (other than the two corner points), we keep the decoding algorithm essentially the same as that for a point-to-point system. This low complexity design makes it attractive for practical implementations.

We should also point out that although this paper exclusively considers CFMA with LDPC codes, the same methodology can be applied to any linear channel codes (e.g., convolutional codes) with appropriate modifications.

\subsection{Code construction and encoding}
\label{subsec:basicCFMA_code_constr}

Let $(R_1,R_2)$ be the target rate pair ($R_1,R_2\leq 1$ for  binary codes) and assume $R_1\geq R_2$ w.l.o.g. In principle, we need to find two good channel codes $\mathcal{C}_1, \mathcal{C}_2$ with rates $R_1, R_2$ and satisfying $\mathcal{C}_2\subseteq\mathcal{C}_1$.  For LDPC binary codes, we describe a method to construct two nested codes, given that we already have one binary LDPC code.

\textbf{Code construction with binary LDPC codes}: Assume $R_1\geq R_2$, we pick an LDPC code $\mathcal{C}_2$ of rate $R_2$ for user $2$, with its parity check matrix $\ve{H}\in\mathbb F_2^{(n-k_2)\times n}$. To construct the code $\mathcal{C}_1$ for user 1 while ensuring $\mathcal{C}_2\subseteq\mathcal{C}_1$, a ``merging" technique is used as shown below. For example, let $\ve{h}_i^\mathsf{T},\ve{h}_j^\mathsf{T}\in\mathbb F_2^{1\times n}$ be two rows of the parity-check matrix $\ve{H}$ of the code $\mathcal{C}_2$. Since any codeword $\ve{u}$  from $\mathcal{C}_2$ satisfies $\ve{h}_i^\mathsf{T}\ve{u}=0,\quad \ve{h}_j^\mathsf{T}\ve{u}=0,$ it also satisfies $(\ve{h}_i\oplus \ve{h}_j)^\mathsf{T}\ve{u}=0.$
Replacing two rows $\ve{h}_i^\mathsf{T}, \ve{h}_j^\mathsf{T}$ in $\ve{H}$ by the new row $(\ve{h}_i\oplus \ve{h}_j)^\mathsf{T}$ we obtain a new code $\mathcal C'$. The parity check matrix $\ve{H}'$ of $\mathcal C'$ is of dimension $(n-k_2-1)\times n$ hence has a higher rate. Obviously, any codeword $\ve{u}\in\mathcal{C}_2$ satisfies $\ve{H}'\ve{u}=\ve{0}$, hence is a codeword of $\mathcal C'$. Equivalently, this merging procedure can also be represented using the Tanner graph of the LDPC code, as shown in Figure~\ref{fig:merging} where two check nodes are merged into one. By repeating this procedure, we can merge more and more rows in the parity check matrix  of one LDPC code to obtain a new LDPC code, with the property that the former code is contained in  the latter.

\begin{example}
We give an example of constructing two nested LDPC codes in Figure~\ref{fig:merging} by merging check nodes. The original LDPC code is shown in Figure~\ref{fig:1a} with four check nodes $f_1,f_2,f_3,f_4$  where check nodes $f_3$ and $f_4$ impose the constraints:
\begin{align*}
x_3\oplus x_5 \oplus x_6 \oplus x_8=0, \quad x_4\oplus x_5\oplus x_6 \oplus x_7=0
\end{align*}
We merge the check nodes $f_3$ and $f_4$ to obtain a new code in Figure~\ref{fig:1b} with three check nodes $f_1',f_2',f_3'$. Since $f_3'$ is formed by merging $f_3$ and $f_4$, it imposes the constraint:
\begin{align*}
x_3\oplus x_4 \oplus x_7 \oplus x_8=0
\end{align*}
The check nodes $f_1'$ and $f_2'$ give the same constraints as $f_1$ and $f_2$ respectively. The rate of the new code is increased to $5/8$ from the original code with a rate $1/2$.

\begin{remark}
A problem which could potentially rise from this merging process is that, some variable nodes can be left isolated after two check nodes are merged. For example consider a code where the variable node $x_1$ is only  connected to two check nodes as 
\begin{align*}
x_1\oplus x_2\oplus x_3=0, \quad x_1\oplus x_4\oplus x_5=0.
\end{align*}
After merging these two check nodes, a new check node is formed to give the constraint $x_2\oplus x_3\oplus x_4\oplus x_5=0$ and the variable node $x_1$ is left isolated because it was not connected to any other check node. This is a situation we want to avoid in merging check nodes. A sufficient condition is that we will only merge two check nodes, if they have disjoint neighbors. This condition is not as stringent as it seems for LDPC codes, and is satisfied for most check nodes due to the sparsity of the code. Also notice that if we merge check nodes using a  LDPC code where all check nodes have odd degree, this issue will not arise.
\end{remark}

\tikzstyle{vertex}=[circle, draw, inner sep=0pt, minimum size=11pt]


\begin{figure}[ht]%
\centering
\subfloat[Original Tanner graph]{
	\begin{tikzpicture}[every path/.append style={thick}, every node/.style={draw,minimum size=3mm}]
\foreach \i in {1,...,8} {
	\node[shape=circle] (v\i) at (-2.25+\i*.5,0) [label=below:$x_{\i}$] {};
};
\foreach \i in {1,...,4} {
	\node (c\i) at (-1.5+\i*.6,1.5) [label=above:$f_{\i}$] {};
};
\path
	(c1) edge (v1)
	(c1) edge (v6)
	(c1) edge (v7)
	(c1) edge (v8)
	(c2) edge (v2)
	(c2) edge (v5)
	(c2) edge (v7)
	(c2) edge (v8)
	(c3) edge (v3)
	(c3) edge[red] (v5)
	(c3) edge[red] (v6)
	(c3) edge (v8)
	(c4) edge (v4)
	(c4) edge[red] (v5)
	(c4) edge[red] (v6)
	(c4) edge (v7)
;
\end{tikzpicture}
	\label{fig:1a}
}%
\subfloat[Tanner graph after merging]{
	\begin{tikzpicture}[every path/.append style={thick}, every node/.style={draw,minimum size=3mm}]
\foreach \i in {1,...,8} {
	\node[shape=circle] (v\i) at (-2.25+\i*.5,0) [label=below:$x_{\i}$] {};
};
\foreach \i in {1,2,3} {
	\node (c\i) at (-1.2+\i*.6,1.5) [label=above:$f'_{\i}$] {};
};
\path
	(c1) edge (v1)
	(c1) edge (v6)
	(c1) edge (v7)
	(c1) edge (v8)
	(c2) edge (v2)
	(c2) edge (v5)
	(c2) edge (v7)
	(c2) edge (v8)
	(c3) edge (v3)
	(c3) edge (v4)
	(c3) edge (v7)
	(c3) edge (v8)
;
\end{tikzpicture}
	\label{fig:1b}
}\\
\subfloat[Original parity-check matrix]{
\begin{tikzpicture}
\node {$\small
	\ve{H}
    = \arraycolsep=1mm \def\arraystretch{1.2}
    \begin{bmatrix}
		1 & 0 & 0 & 0 & 0 & 1 & 1 & 1 \\
		0 & 1 & 0 & 0 & 1 & 0 & 1 & 1 \\
		\tikzmark{left}{0} & 0 & 1 & 0 & \tikzmark{merge1top}{1} & \tikzmark{merge2top}{1} & 0 & 1 \\
		0 & 0 & 0 & 1 & \tikzmark{merge1bottom}{1} & \tikzmark{merge2bottom}{1} & 1 & \tikzmark{right}{0}
    \end{bmatrix}$
};
\end{tikzpicture}
\label{fig:1c}
}%
\subfloat[Parity-check matrix after merging]{
\begin{tikzpicture}
\node {$\small
	\ve{H}
    = \arraycolsep=1mm \def\arraystretch{1.2}
    \begin{bmatrix}
		1 & 0 & 0 & 0 & 0 & 1 & 1 & 1 \\
		0 & 1 & 0 & 0 & 1 & 0 & 1 & 1 \\
		\tikzmark{mergedleft}{0} & 0 & 1 & 1 & \tikzmark{merged1}{0} & \tikzmark{merged2}{0} & 1 & \tikzmark{mergedright}{1}
    \end{bmatrix}$
};
\end{tikzpicture}
\label{fig:1d}
}%
\caption{How to construct nested linear codes by parity-check merging. Tanner graph and parity-check matrix of the original code \protect\subref{fig:1a}, \protect\subref{fig:1c}; of the derived supercode \protect\subref{fig:1b}, \protect\subref{fig:1d}.}
\label{fig:merging}
\end{figure}
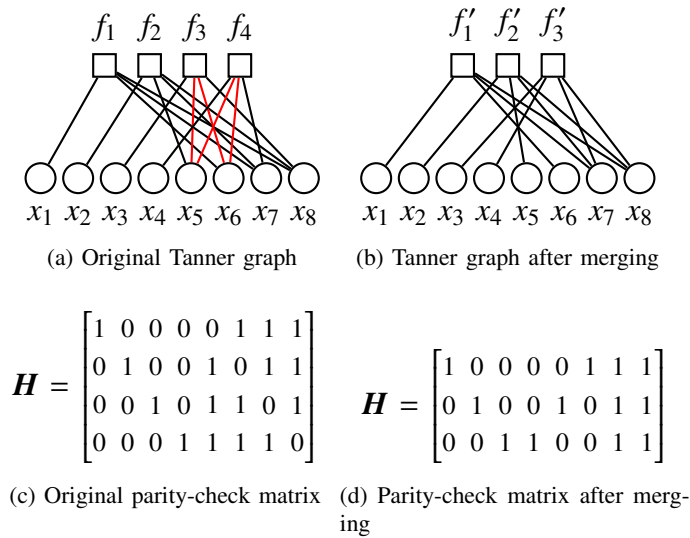
\end{example}

\textbf{Encoding and modulation}: Given two messages $w_1, w_2$ from two users, the codewords $\ve{u}_1,\ve{u}_2$ are generated using nested codebooks $\mathcal{C}_1, \mathcal{C}_2$. The binary codewords are mapped to the real-valued channel input using the BPSK mapping where every bit is mapped to one symbol. 
In particular, we have for $i=1,\ldots, n$
	$(x_{1,i},x_{2,i})
	= (\varphi_1(u_{1,i}),\varphi_2(u_{2,i}))$
where $\varphi_k \colon \mathbb{F}_2 \to \mathcal{X}$ is defined as $\varphi_k(u_{k,i}) = \sqrt{P}(2 u_{k,i} - 1)$ for $k={1,2}$.

\def\QAMpoint#1{
		\draw[shift={#1}] (.1,.1)--(-.1,-.1);
		\draw[shift={#1}] (-.1,.1)--(.1,-.1);
}

\subsection{Decoding algorithm}
\label{sec:3b}
Now we derive the decoding algorithm for the CFMA scheme using binary LDPC codes, and show that the same sum-product algorithm for the point-to-point LDPC decoding can be directly applied to our scheme with only a slight modification to the initialization step. We define $\ve{s}=\ve{u}_1\oplus \ve{u}_2$, and derive the algorithm for decoding the pair $(\ve{s}, \ve{u}_1)$. The decoding procedure for the pair $(\ve{s}, \ve{u}_2)$ is similar.


The decoder uses a bit-wise maximum {\em a posteriori} (MAP) estimation for decoding, i.e.
\begin{itemize}
\item{\makebox[2cm]{Decode $\ve{s}$: \hfill} ${\hat{s}}_{i}=\argmax\limits_{s_{i}}p(s_i|\ve{y})$}
\item{\makebox[2cm]{Decode $\ve{u}_1$: \hfill} ${\hat{u}}_{1,i}=\argmax\limits_{u_{1,i}}p(u_{1,i}|\ve{y},\ve{s})$}
\end{itemize}

In the following derivation we use $\tilde{\mathcal C}$ to  denote the codes with the larger rate among $\mathcal{C}_1, \mathcal{C}_2$ and  $\tilde{\ve{H}}$ to denote its parity check matrix. Ideally, we target a bit-wise maximum {\em a posteriori} (MAP) estimation, i.e., $\argmax_{s_i \in \{0,1\}} p(s_i| \ve{y})$. However, since $p(\yv|\sv)$ is not memoryless, the sum-product algorithm does not directly approximate the bit-wise MAP in this case. Nonetheless, as an approximation to the bit-wise MAP rule, we perform a bit-wise MAP estimation as follows:
\begin{IEEEeqnarray*}{rCl}
	\hat{s}_i &=& \argmax_{s_i \in \{0,1\}} \sum_{\sim s_{i}} \prod_{i=1}^{n} p(y_{i}|s_{i}) \mathbbm{1}\bigl\{ \tilde{\ve{H}}\ve{s} = \ve{0} \bigr\}, \IEEEyesnumber\label{eq:qnst1}
\end{IEEEeqnarray*}
where the summation is over all coordinates of $\ve{s}$ except $s_i$. 
We also use the fact that $\ve{s} = \ve{u}_1 \oplus \ve{u}_2$ is uniformly distributed over $\tilde{\mathcal{C}}$ as a consequence of $\ve{u}_1$ and $\ve{u}_2$ being uniform over the  nested codebooks $\mathcal{C}_1$ and $\mathcal{C}_2$, respectively, hence
\begin{align*}
	p(\ve{s})
	 = \frac{\mathbbm{1}\bigl\{ \tilde{\ve{H}}\ve{s} = \ve{0} \bigr\}}{|\tilde{\mathcal C}|}.
\end{align*}
As shown in \cite{Richardson--Urbanke2008}, the formulation in \eqref{eq:qnst1} has complexity in the order of standard sum-product algorithm for the bit-wise MAP estimation of the sum codewords $\ve{s}$. Also notice that the complexity of this algorithm is the same as for a point-to-point system where the receiver decodes one codeword from the code described by $\tilde{\ve{H}}$.

Similarly, {\em given the sum codeword} $\ve{s} = \ve{u}_1 \oplus \ve{u}_2 \in \tilde{\mathcal{C}}$, we can rewrite the second decoding step as
\begin{IEEEeqnarray*}{rCl}
	\hat{u}_{1,i} &=& \argmax_{u_{1,i} \in \left\{0,1\right\}} p(u_{1,i}|\ve{y},\ve{s}) = \argmax_{u_{1,i} \in \left\{0,1\right\}} \sum_{\sim u_{1,i}}p(\ve{y}|\ve{u}_{1},\ve{s})p(\ve{u}_1,\ve{s}) \\
	&=& \argmax_{u_{1,i} \in \left\{0,1\right\}} \sum_{\sim u_{1,i}} \prod_{i=1}^{n} p(y_{i}|u_{1,i},s_{i}) \mathbbm{1}\bigl\{ \ve{H}_1\ve{u}_1 = \ve{0} \bigr\}.   \IEEEeqnarraynumspace\IEEEyesnumber\label{eq:qnst2}
\end{IEEEeqnarray*}
For the last equality we have used the fact that the channel is memoryless, as well as the fact that $(\ve{u}_1,\ve{s})$ is uniform over $\mathcal{C}_1 \times \tilde{\mathcal{C}}$, hence
\begin{IEEEeqnarray}{rCl} \label{eq:qn3}
	p(\ve{u}_1,\ve{s})
	 &=& \frac{ \mathbbm{1}\bigl\{ \ve{u}_1 \in \mathcal{C}_1 \bigr\} \mathbbm{1}\bigl\{ \ve{s} \in \tilde{\mathcal{C}} \bigr\} }{ |\mathcal{C}_1|\,|\tilde{\mathcal{C}}|},
\end{IEEEeqnarray}
where we recall that  $\ve s$ is the decoded  codeword from $\tilde{\mathcal C}$ hence it always holds that $\ve s\in\tilde{\mathcal C}$, namely $\mathbbm{1}\bigl\{ \ve{s} \in \tilde{\mathcal{C}} \bigr\} =1$. Furthermore $\ve{u}_1 \in \mathcal{C}_1$ is equivalent to $\ve{H}_1\ve{u}_1 = \ve{0}$.



It is important to realize that Eq.~\eqref{eq:qnst2} also takes the same form as Eq.~\eqref{eq:qnst1}, hence allow us to carry out the optimization efficiently using the same sum-product algorithm. More precisely, both decoding steps in \eqref{eq:qnst1} and \eqref{eq:qnst2} can be implemented using the standard sum-product algorithm used in the LDPC decoder for a point-to-point system, with a modified initial log-likelihood ratio (LLR) values (derivations in Appendix \ref{appen:derivations_LLR_basic}): 
\begin{subequations}
\begin{IEEEeqnarray}{rRl}
	\LLR_1
	&\coloneqq& \log \frac{p(y_i|s_i=0)}{p(y_i|s_i=1)} = \log\cosh\bigl(y_i 2\sqrt{P}\bigr) - 2P \\
	\LLR_2
	 &\coloneqq& \log \frac{p(y_i|u_{1,i}=0,s_i)}{p(y_i|u_{1,i}=1,s_i)} =
	 	\begin{cases}
			4 y_i\sqrt{P} & \text{for $s_i=0$} \\
			0 & \text{for $s_i=1$}
		\end{cases}
\end{IEEEeqnarray}
\label{eq:LLR_basic_CFMA}
\end{subequations}

Algorithm~\ref{alg:cfma} highlights the decoding process for the basic CFMA scheme with binary LDPC codes.  The function SPA$(\ve H, \LLR)$ executes the standard sum-product algorithm on the Tanner graph given by the parity-check matrix $\ve H$ with initial input value $\LLR$. Details and efficient implementation of this standard algorithm can be found in many existing literature, e.g. \cite{Hu--Eleftheriou--Arnold--Dholakia2001}. 
\begin{algorithm}[h!]
\caption{CFMA: Decoding algorithm with binary LDPC codes. $\LLR_1$ and $\LLR_2$ are given in (\ref{eq:LLR_basic_CFMA}).}
\begin{algorithmic}[1]
\State $\hat{\ve{s}} = \mathrm{SPA}(\tilde{\ve{H}}, \LLR_1$) \Comment{Decode the sum codeword $\ve{s}$}
\State $\hat{\ve{u}}_1 = \mathrm{SPA}(\ve{H}_1, \LLR_2$)\Comment{Decode the codeword $\ve{u}_1$}
\State $\hat{\ve{u}}_2 = \hat{\ve{s}} \oplus \hat{\ve{u}}_1$  \Comment{Decode the codeword $\ve{u}_2$}
\end{algorithmic}
\label{alg:cfma}
\end{algorithm}

\section{CFMA with multilevel binary codes} \label{sec:CFMA extensions}

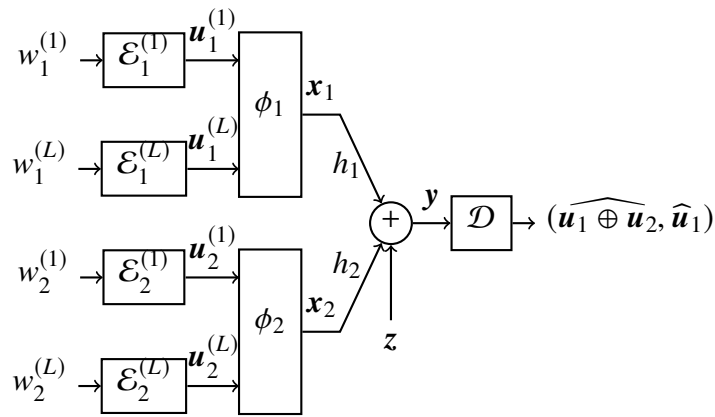
\begin{figure}[ht]
\begin{center}
\begin{tikzpicture}[every node/.style={anchor=base, text height=.5em, text depth=.25em}, every path/.append style={thick}]
\matrix[matrix of math nodes, row sep=7mm, column sep=7mm, inner sep=2mm, nodes in empty cells, nodes={minimum width=8mm}] (modenc) {
	|[draw]| \mathcal{E}_{1}^{(1)} & \\
	|[draw]| \mathcal{E}_{1}^{(L)} & \\
	|[draw]| \mathcal{E}_{2}^{(1)} & \\
	|[draw]| \mathcal{E}_{2}^{(L)} & \\
};
\node[draw, inner sep=0, fit=(modenc-1-2) (modenc-2-2)] (M1) {$\phi_1$};
\node[draw, inner sep=0, fit=(modenc-3-2) (modenc-4-2)] (M2) {$\phi_2$};
\node[draw,shape=circle,inner sep=.5mm,right=7mm of modenc] (adder) {$+$};
\draw[<-] (modenc-1-1.west)--+(-3mm,0) node[left] {$w_{1}^{(1)}$};
\draw[<-] (modenc-2-1.west)--+(-3mm,0) node[left] {$w_{1}^{(L)}$};
\draw[<-] (modenc-3-1.west)--+(-3mm,0) node[left] {$w_{2}^{(1)}$};
\draw[<-] (modenc-4-1.west)--+(-3mm,0) node[left] {$w_{2}^{(L)}$};
\draw[->] (modenc-1-1.east)--(modenc-1-2.west) node[midway,above] {$\ve{u}_{1}^{(1)}$};
\draw[->] (modenc-2-1.east)--(modenc-2-2.west) node[midway,above] {$\ve{u}_{1}^{(L)}$};
\draw[->] (modenc-3-1.east)--(modenc-3-2.west) node[midway,above] {$\ve{u}_{2}^{(1)}$};
\draw[->] (modenc-4-1.east)--(modenc-4-2.west) node[midway,above] {$\ve{u}_{2}^{(L)}$};
\draw[->] (M1.east)--+(5mm,0) node[midway,above] {$\ve{x}_1$} -- (adder) node[midway,xshift=-1ex,yshift=-1ex] {$h_1$};
\draw[->] (M2.east)--+(5mm,0) node[midway,above] {$\ve{x}_2$} -- (adder) node[midway,xshift=-1ex,yshift=1ex] {$h_2$};
\draw[<-] (adder.south)--+(0,-1cm) node[below] {$\ve{z}$};
\node[draw, right=5mm of adder, inner sep=2mm] (dec) {$\mathcal{D}$};
\draw[->] (adder.east)--(dec.west) node[midway,above] {$\ve{y}$};
\draw[->] (dec.east)--+(3mm,0) node[right] {$(\widehat{\ve{u}_1 \oplus \ve{u}_2},\widehat{\ve{u}}_1)$};
\end{tikzpicture}
\caption{Block diagram for CFMA with multilevel binary codes. The codewords $\ve u_1, \ve u_2$ are constructed using multiple binary codes $\ve u_1^{\ell}, \ve u_2^{\ell}, \ell=1,\ldots, L$. } \label{fig:ll}
\end{center}
\end{figure}
The construction in the previous section with BPSK modulation can only achieve a communication rate up to $1$ bit per dimension. In this section, we present how to extend our approach to support higher rates.
To support rates higher than 1 bit per real dimension, a direct approach would be to use non-binary codes with PAM modulation in which the number of symbols equals the alphabet size of the codes. However, such approach would require the construction of complex non-binary codes and a decoding algorithms to handle non-binary symbols. Instead, as an alternative to this approach we present a multilevel CFMA strategy to support higher rates based on {\em binary} codes. The use of binary codes in a multilevel fashion will allow the practical CFMA strategy to be more consistent and compatible with current practical systems which are mostly based on binary codes in conjunction with PAM modulation.
With the goal to design low-complexity CFMA codes that are also compatible with such architectures, we present a high-order CFMA strategy based on multilevel binary codes \cite{Forney2000} \cite{Wachsmann99}.
We note that by restricting to multilevel binary codes, the proposed strategy is different from directly constructing codes from non-binary fields, and thus we are only loosely inspired by theoretical achievability results in Theorem~\ref{thm:cfma-nlc} for higher rates.

\subsection{Code construction and encoding}

Let $L$ denote the number of code levels. User~1 employs a collection $(\mathcal C_1^{(1)},\ldots,\mathcal C_1^{(L)})$ of binary $(k_1^{(\ell)},n)$-codes, $\ell=1,\dotsc, L$, such that $k_1^{(1)} + \dotso + k_1^{(L)} = k_1$. Similarly, user~2 employs a collection $(\mathcal C_2^{(1)},\ldots,\mathcal C_2^{(L)})$ of binary $(k_2^{(\ell)},n)$-codes, $\ell=1,\ldots, L$, such that $k_2^{(1)} + \dotso + k_2^{(L)} = k_2$. Moreover, we require that the codebooks of level $\ell\in\{1,\ldots, L\}$ be nested between users. For simplicity we assume that $\mathcal C_1^{(\ell)}\subseteq\mathcal C_2^{(\ell)}$ for all $\ell=1,\ldots, L$. These nested codebooks may be constructed using the merging method described in Section~\ref{subsec:basicCFMA_code_constr}. 



\textbf{Encoding and modulation}: For user \num{1}, we split the message $w_1$, of length $k_1$, into $L$ submessages $w_1^{(1)},\ldots,w_1^{(L)}$, of respective lengths $k_1^{(\ell)},\ \ell=1,\dotsc,L$. Each submessage $w_1^{(\ell)}$ is then encoded independently by the corresponding $(k_1^{(\ell)},n)$-binary code into a length-$n$ subcodeword $\ve{u}_1^{(\ell)} \in \{0,1\}^n$, $\ell=1,\ldots,L$.
We proceed likewise for user \num{2}.
Hence, the rates of both users are respectively
	$R_1
    = \frac{k_1}{n}
    = \frac{\sum_{\ell=1}^L k_1^{(\ell)}}{n}$
and
	$R_2= \frac{k_2}{n}
    = \frac{\sum_{\ell=1}^L k_2^{(\ell)}}{n}.$
The bijective modulation mapping 
	$\varphi_1 \colon \{0,1\}^L \to \mathcal{X}$
is a symbol-by-symbol mapping expressible as the composition of two functions: first, for every $i=1,\dotsc,n$, a $2^L$-ary codeword symbol
\begin{equation}
	u_{1,i}
	= \sum_{\ell=1}^L 2^{\ell-1} u_{1,i}^{(\ell)}   \label{eq:u}
\end{equation}
is formed from the binary subcodeword symbols; secondly, the $2^L$-ary symbol is mapped one-to-one to a signal-space symbol $x_{1,i} \in \mathcal{X}$. For simplicity, we will choose the latter to be affine-linear. All in all, the signal-space codeword $\ve{x}_1$ can thus be represented as
\begin{IEEEeqnarray*}{rCl}
	\ve{x}_1
	&=& \varphi_1\bigl(\ve{u}_1^{(1)},\dotsc,\ve{u}_1^{(L)}\bigr) = \sqrt{\frac{3P}{2^{2L}-1}} \sum_{\ell=1}^L 2^{\ell-1} \bigl(2 u_1^{(\ell)}-1\bigr)   \IEEEeqnarraynumspace\IEEEyesnumber\label{x1}
\end{IEEEeqnarray*}
where it is understood that $\varphi_1$ is applied symbolwise. The factor to the left of the summation symbol in \eqref{x1} ensures that the average power constraint is met. The mapping $\varphi_2$ is defined similarly.

\subsection{Decoding algorithm} \label{sec:4b}

We will view codewords $\ve{u}_1$ and $\ve{u}_2$ constructed as in \eqref{eq:u} as vectors in $\mathbb{Z}_{2^L}$, and first decode the sum codeword $\ve{s}$ defined as 
\begin{align}
	\ve{s} \coloneqq [\ve{u}_1 + \ve{u}_2] \mod 2^L
\label{eq:s}
\end{align}
where the sum is performed element-wise between $\ve{u}_1$ and $\ve{u}_2$. Importantly, we will show that the same sum-product algorithm for the point-to-point LDPC decoding can be applied to the proposed scheme as well. 

Since $\ve s$ is an element in $\mathbb Z_{2^L}$, decoding $\ve s$ directly would require an algorithm which can handle symbols in $\mathbb Z_{2^L}$. As we wish to reuse the existing sum-product algorithm for binary codes, we will first represent the sum $\ve s$ in its binary form. That is, each entry $s_i$ is  written as 
$s_i=\sum_{\ell=1}^L2^{\ell-1} s_i^{(\ell)}$
for some $s_i^{\ell}\in\{0,1\}$, for all $i=1,\ldots, n$. 

The following lemma relates the binary representation of $\ve s$ with $\ve u_1,\ve u_2$.
  \begin{lemma}
  Let $\ve u_1, \ve u_2$ be two $n$-length strings in $\mathbb Z_{2^L}^n$ constructed as in \eqref{eq:u} using $\ve u_{1}^{(\ell)},\ve u_{2}^{(\ell)} \text{ for } \ell=1,\ldots, L$. Let $\ve s$ take the form as in \eqref{eq:s}. Then we have the following relationships:
  \begin{enumerate}
  \item $\ve s^{(1)}=\ve u_{1}^{(1)}\oplus_2\ve u_{2}^{(1)}$
  \item For $2\leq \ell\leq L$ and for each $i=1,\ldots, n$ define the partial sum
   $s_{i}^{\ell_+}:=\sum_{j=1}^{\ell-1} 2^{j-1}(u_{1,i}^{(j)}+u_{2,i}^{(j)}).$
Then,
\begin{itemize}
\item if $s_{i}^{\ell_+}<2^{\ell-1}$, we have $ s_{i}^{(\ell)}=u_{1,i}^{(\ell)}\oplus_2 u_{2,i}^{(\ell)}$;
\item if $s_{i}^{\ell_+}\geq 2^{\ell-1}$, we have  $ s_{i}^{(\ell)}= u_{1,i}^{(\ell)}\oplus_2 u_{2,i}^{(\ell)}\oplus_2 1$.
\end{itemize}
\end{enumerate} 
\label{lemma:relationship}
\end{lemma}

\begin{IEEEproof}
To prove 1), notice that for any $i=1,\ldots n$, if $( u_{1,i}^{(1)}, u_{2,i}^{(1)})$ equals to $(1,1)$ or $(0,0)$, then  $ \sum_{\ell=1}^L2^{\ell-1}u^{(\ell)}_{1,i}$ and $\sum_{\ell=1}^L 2^{\ell-1}u^{(\ell)}_{2,i}$ are either both odd numbers or both even numbers, for $i=1,\ldots, n$. As a result $\sum_{\ell=1}^L2^{\ell-1} s^{(\ell)}_i$ is an even number hence $ s_{i}^{(1)}= u_{1,i}^{(1)}\oplus_2 u_{2,i}^{(1)}=0$ for all $i=1,\ldots, n$. On the other hand if we have $ u_{1,i}^{(1)}+ u_{2,i}^{(1)}=1$, then $\sum_{\ell=1}^L2^{\ell-1} s^{(\ell)}_i$ is an odd number hence $ s_{i}^{(1)}=1$.

Now we consider $\ve s_{i}^{(\ell)}$ for $\ell\geq 2$, in which case the relationship between $s_{i}^{(\ell)}$ and $( u_{1,i}^{(\ell)},u_{2,i}^{(\ell)})$ is more complicated because of the carry over from the lower digits. First notice that we have
\begin{align*}
s_i=&2^{L-1}(u_{1,i}^{(L)}+u_{2,i}^{(L)})+\cdots+2(u_{1,i}^{(2)}+u_{2,i}^{(2)})+(u_{1,i}^{(1)}+u_{2,i}^{(1)}) +d_i\cdot 2^L
\end{align*}
where $d_i$ equals either $0$ or $-1$. Meanwhile we also have $s_i=2^{L-1} s_{i}^{(L)}+\ldots+2 s_{i}^{(2)}+ s_{i}^{(1)}.$
For a given $\ell\geq 2$, it is easy to see that if the partial sum $s_{i}^{\ell_+}$ satisfies $s_{i}^{\ell_+}<2^{\ell-1}$, then we have
$s_{i}^{\ell_+}=\sum_{j=1}^{\ell-1} 2^{j-1} s_{i}^{(j)}.$
In this case $ s_{i}^{(\ell)}$ will not be affected by the carry over from the lower digits $s_{i}^{(1)},\ldots, s_{i}^{(\ell-1)}$ and is solely determined by $u_{1,i}^{(\ell)}, u_{2,i}^{(\ell)}$. Particularly, it is straightforwardly to check that $s_{i}^{(\ell)}=0$ if $(u_{1,i}^{(\ell)}, u_{2,i}^{(\ell)})$ equals $(0,0)$ or $(1,1)$ (in the latter case there is a carry over to $\ve s_{i}^{(\ell+1)}$), and $s_{i}^{(\ell)}=1$ if $(u_{1,i}^{(\ell)}, u_{2,i}^{(\ell)})$ equals $(0,1)$ or $(1,0)$. Namely we have $s_{i}^{(\ell)}= u_{1,i}^{(\ell)}\oplus_2 u_{2,i}^{(\ell)}$ in this case.

If the partial sum $s_{i}^{\ell_+}$ satisfies $s_{i}^{\ell_+}\geq 2^{\ell-1}$, then we have
$s_{i}^{\ell_+}=\sum_{j=1}^{\ell-1} 2^{j-1} s_{i}^{(j)} +  2^{\ell-1}.$
In this case $s_{i}^{(\ell)}$ is determined by $u_{1,i}^{(\ell)}, u_{2,i}^{(\ell)}$ as well as the carry over from the lower digits. It is also straightforward to check that in this case we have $s_{i}^{(\ell)}=0$ if $(u_{1,i}^{(\ell)}, u_{2,i}^{(\ell)})$ equals $(0,1)$ or $(1,0)$, $s_{i}^{(\ell)}=1$ if $( u_{1,i}^{(\ell)},  u_{2,i}^{(\ell)})$ equals $(0,0)$, and $s_{i}^{(\ell)}=1$ if $( u_{1,i}^{(\ell)},  u_{2,i}^{(\ell)})$ equals $(1,1)$  because of the carry over. Namely we have $s_{i}^{(\ell)}= u_{1,i}^{(\ell)}\oplus_2  u_{2,i}^{(\ell)}\oplus_2 1$ in this case.
\end{IEEEproof}

The above lemma helps us design the appropriate decoding algorithm of this construction using existing sum-product algorithm. Briefly, our decoding algorithm will decode the codewords $(\ve s, \ve u_1)$ in consecutive steps, in particular the decoder will output the codewords in the order: $\ve s^{(1)}, 
\ve  u_{1}^{(1)}, \ve s^{(2)}, \ve u_{1}^{(2)},\ldots, \ve s^{(L)}, \ve u_{1}^{(L)}$.

Recall that the codes $\mathcal C_{1}^{(\ell)},\mathcal C_{2}^{(\ell)}$ are nested for all $\ell=1,\ldots, L$. Let $\tilde{\mathcal C}^{(\ell)}$ denote one of the two codes with the larger rate, and $\tilde{\ve H}^{(\ell)}$ denote its corresponding parity check matrix. We start our derivation for  decoding the bit-string $\ve s^{(1)}$, i.e., the string of the least significant bits in the binary representation of the sum $\ve s$. Similar to the single-layer case in the previous section, we perform the following estimation as an approximation to the bit-wise MAP rule for $i=1,\ldots, n$:
\begin{align}
{\hat{s}}_{i}^{(1)} &=\argmax_{s_i^{(1)} \in \{0,1\}} \sum_{\sim s_{i}^{(1)}} \prod_{i=1}^{n} p(y_{i}|s_{i}^{(1)})p(\ve s^{(1)}) \nonumber \\
&\overset{(a)}{=} \argmax_{s_{i}^{(1)} \in \left\{0,1\right\}} \sum_{\sim s_{i}^{(1)}} \prod_{i=1}^{n}  p(y_{i}|s_{i}^{(1)}) \mathbbm{1} \bigl\{ \tilde{\ve H}^{(1)}\ve s^{(1)}=\ve{0} \bigr\}
\end{align}
where step $(a)$ follows from Lemma~\ref{lemma:relationship} and that we have $\ve s^{(1)}=\ve u_{1}^{(1)}\oplus\ve u_{2}^{(1)}$. This ensures that $\ve s^{(1)}$ is a codeword from $\tilde{\mathcal C}^{(1)}$, hence
$p(\ve{s}^{(1)}) = \frac{\mathbbm{1}\bigl\{ \tilde{\ve{H}}^{(1)}\ve s^{(1)} = \ve{0} \bigr\}}{|\tilde{\mathcal C}^{(1)}|}.$

As for the input for the sum-product algorithm, the initial LLR value is given by
\begin{align} \label{eqn:LLRmult1}
&\LLR_1^{(1)} =  \log \frac{p(y_i|s_i^{(1)}=0)}{p(y_i|s_i^{(1)}=1)}
\end{align}
where the explicit expression is given in Appendix \ref{appen:derivations_LLR_multilevel}.

With the above expression, the bit string $\ve s^{(1)}$ can readily be decoded using the sum-product algorithm for binary codes. Next we decode $\ve u_{1}^{(1)}$ (or $\ve u_{2}^{(1)}$) using channel output $\ve y$ and the decoded bit string $\ve s^{(1)}$. For $i=1,\ldots,n$
\begin{align}
{\hat{u}}_{1,i}^{(1)}&=\argmax_{u_{1,i}^{(1)} \in \left\{0,1\right\}}p(u_{1,i}^{(1)}|\ve y,\ve s^{(1)}) \nonumber \\
&=\argmax_{u_{1,i}^{(1)} \in \left\{0,1\right\}} \sum_{\sim u_{1,i}^{(1)}}p(\ve y|\ve u_{1}^{(1)},\ve s^{(1)})p(\ve s^{(1)},\ve u_{1}^{(1)}) \nonumber \\
&\overset{(b)}{=}\argmax_{u_{1,i}^{(1)} \in \left\{0,1\right\}} \sum_{\sim u_{1,i}^{(1)}} \prod_{i=1}^{n}  p(y_{i}|s_{i}^{(1)},u_{1,i}^{(1)}) \mathbbm{1} {\bigl\{ {\ve H_{1}^{(1)}}\ve u_{1}^{(1)}=\ve{0} \bigr \}}
\label{eq:u1_multilevel_level_1}
\end{align}
whereas $(b)$ follows from the similar argument as in (\ref{eq:qn3}), namely
	$p(\ve s^{(1)},\ve u_1^{(1)})
	 = \frac{ \mathbbm{1}\bigl\{ \ve{s}^{(1)} \in \tilde{\mathcal{C}}^{(1)} \bigr\} \mathbbm{1}\bigl\{ \ve{u}_1^{(1)} \in \mathcal{C}_1^{(1)} \bigr\} }{|\tilde{\mathcal{C}}^{(1)}| |\mathcal{C}_1^{(1)}|}.$
Note that $\mathbbm{1}\bigl\{ \ve{s}^{(1)} \in \tilde{\mathcal{C}}^{(1)} \bigr\}=1$ in this case because $\ve s^{(1)}$ is the decoded codeword in $\tilde{\mathcal C}^{(1)}$. Furthermore $\ve{u}_1^{(1)} \in \mathcal{C}_1^{(1)}$ is equivalent to $\ve{H}_1^{(1)}\ve{u}_1^{(1)} = \ve{0}$. Also notice that at this point we have $\ve s^{(1)}, \ve u_{1}^{(1)}$, and thus, we can reconstruct $\ve u_{2}^{(1)}=\ve u_{1}^{(1)}\oplus_2\ve s^{(1)}$ from Lemma~\ref{lemma:relationship}. 

The initial LLR value for the sum-product algorithm is given by
$\LLR_2^{(1)} = \log \frac{p(y_i|s_i^{(1)}, u_{1,i}^{(1)}=0)}{p(y_i|s_i^{(1)}, u_{1,i}^{(1)}=1)}$
with the exact expression given by (\ref{eqn:LLR21}) in Appendix \ref{appen:derivations_LLR_multilevel}. Now we start to decode the $2$-nd level codes $\ve u_{1}^{(2)},\ve u_{2}^{(2)}$, given the decoded information $\ve u_{1}^{(1)},\ve u_{2}^{(1)}$ and the channel output $\ve y$.  To first decode the bit string $\ve s^{(2)}$ from the sum $\ve s$, we use the same MAP approximation for the bit-wise estimation
\begin{align}
{\hat{s}}_{i}^{(2)} &=\argmax_{s_{i}^{(2)} \in \left\{0,1\right\}}p(s_{i}^{(2)}|\ve y,\ve u_{2}^{(1)},\ve u_{1}^{(1)}) \nonumber \\
&= \argmax_{s_{i}^{(2)} \in \left\{0,1\right\}} \sum_{\sim s_{i}^{(2)}}p(\ve y|\ve s^{(2)},\ve u_{2}^{(1)},\ve u_{1}^{(1)})p(\ve s^{(2)}|\ve u_{2}^{(1)},\ve u_{1}^{(1)})p(\ve u_{1}^{(1)},\ve u_{2}^{(1)}) \nonumber \\
&\overset{(c)}{=} \argmax_{s_{i}^{(2)} \in \left\{0,1\right\}} \sum_{\sim s_{i}^{(2)}} \prod_{i=1}^{n} p(y_{i}| s_{i}^{(2)}, u_{2,i}^{(1)},u_{1,i}^{(1)}) \cdot \mathbbm{1}\bigl\{ \tilde{\ve H}^{(2)}(\ve s^{(2)}\oplus_2 \ve c^{(2)})=\ve 0 \bigr\}
\label{eq: s_2}
\end{align}
where $(c)$ follows from the following argument. First observe that $(\ve{u}_1^{(1)},\ve{u}_2^{(1)})$ is uniform over $\mathcal{C}_1^{(1)} \times \mathcal{C}_2^{(1)}$, hence
$p(\ve{u}_1^{(1)},\ve{u}_2^{(1)})
	 = \frac{ \mathbbm{1}\bigl\{ \ve{u}_1^{(1)} \in \mathcal{C}_1^{(1)} \bigr\} \mathbbm{1}\bigl\{ \ve{u}_2^{(1)} \in \mathcal{C}_2^{(1)} \bigr\} }{ |\mathcal{C}_1^{(1)}|\,|\mathcal{C}_2^{(1)}|}$
where $\ve u_{1}^{(1)}, \ve u_{2}^{(1)}$ are guaranteed to be codewords as they are decoded from the previous step. Notice that we already have $\ve u_{1}^{(1)},\ve u_{2}^{(1)}$ hence can construct another $n$-length binary vector $\ve c^{(2)}$ defined as follows
\begin{align*}
c_{i}^{(2)}=
\begin{cases}
0, &\text{if } u_{1,i}^{(1)}+ u_{2,i}^{(1)}< 2\\
1, &\text{if } u_{1,i}^{(1)}+ u_{2,i}^{(1)}=2
\end{cases}
\end{align*}
for all $i=1,\ldots, n$. The value $ c_{i}^{(2)}$ in this vector indicates if there is carry over for $ s_{i}^{(2)}$ from the lower digit. Using the result in Lemma~\ref{lemma:relationship}, we have
\begin{align}
\ve s^{(2)}=\ve u_{1}^{(2)}\oplus_2\ve u_{2}^{(2)}\oplus_2 \ve c^{(2)}
\label{eq:s2}
\end{align}

As $\ve c^{(2)}$ is completely determined by $\ve u_{1}^{(1)},\ve u_{2}^{(1)}$, the term $p(\ve s^{(2)}|\ve u_{2}^{(1)},\ve u_{1}^{(1)})$ can be further developed as
\begin{align}
p(\ve s^{(2)}|\ve u_{2}^{(1)},\ve u_{1}^{(1)})&=p({\ve s^{(2)}|\ve u_{2}^{(1)},\ve u_{1}^{(1)},\ve c^{(2)}} ) \nonumber \\
&\overset{(d)}{=}p(S^{(2)}=\ve s^{(2)}|C^{(2)}=\ve c^{(2)} ) \nonumber \\
&\overset{(e)}{=}p(U_1^{(2)}\oplus_2 U_2^{(2)}=\ve s^{(2)}\oplus_2\ve c^{(2)}|C^{(2)}=\ve c^{(2)} ) \nonumber \\
&\overset{(f)}{=}p(U_1^{(2)}\oplus_2 U_2^{(2)}=\ve s^{(2)}\oplus_2 \ve c^{(2)} ) \nonumber \\
&\overset{(g)}{=}\frac{1}{|\tilde{\mathcal C}^{(2)}|}\mathbbm{1}\bigl\{ \tilde{\ve H}^{(2)}(\ve s^{(2)}\oplus_2 \ve c^{(2)})=\ve 0 \bigr\}
\label{eq:ps_multiplevel_level_2}
\end{align}
where $(d)$ follows since $ \ve s^{(2)}$ and $(\ve u_{1}^{(1)}, \ve u_{2}^{(1)})$ are conditionally independent  given $\ve c^{(2)}$. Step $(e)$ follows because of \eqref{eq:s2} and step $(f)$ follows because $\ve c^{(2)}$ is a function of $\ve u_{1}^{(1)}, \ve u_{2} ^{(1)}$ which are independent from $ \ve u_{1}^{(2)}$ and $\ve u_{2}^{(2)}$. The last step holds because $\ve u_{1}^{(2)}, \ve u_{2}^{(2)}$ are from nested codebooks $\mathcal C_{1}^{(2)}$ and $\mathcal C_{2}^{(2)}$. This completes the argument for step $(c)$.
Also notice that the expression (\ref{eq: s_2}) already admits a sum-product algorithm for decoding the sequence $\ve s^{(2)}$.
Furthermore, the corresponding initial LLR value is given by
\begin{align}
\LLR_1^{(2)}&=\log \frac{p(y_i| u_{1,i}^{(1)}, u_{2,i}^{(1)}, s_{i}^{(2)} \oplus c_i^{(2)}=0)}{p(y_i|  u_{1,i}^{(1)}, u_{2,i}^{(1)}, s_{i}^{(2)} \oplus c_i^{(2)}=1)}
\end{align}
with an explicit expression given by (\ref{eqn:LLR12}) Appendix \ref{appen:derivations_LLR_multilevel}. The next step is to decode  $\ve u_{1}^{(2)}$:
\begin{align}
{\hat{u}}_{1,i}^{(2)}&=\argmax_{u_{1,i}^{(2)} \in \left\{0,1\right\}}p(u_{1,i}^{(2)}|\ve y,\ve s^{(2)},\ve u_{1}^{(1)},\ve u_{2}^{(1)}) \nonumber \\
&=\argmax_{u_{1,i}^{(2)} \in \left\{0,1\right\}} \sum_{\sim u_{1,i}^{(2)}}p(\ve y|\ve u_{1}^{(1)},\ve u_{2}^{(1)},\ve s^{(2)}, \ve u_{1}^{(2)}) \cdot p(\ve s^{(2)},\ve u_{1}^{(2)},\ve u_{1}^{(1)},\ve u_{2}^{(1)}) \nonumber \\
&\overset{(h)}{=}\argmax_{u_{1,i}^{(2)} \in \left\{0,1\right\}} \sum_{\sim u_{1,i}^{(2)}} \prod_{i=1}^{n} p(y_{i}|u_{1,i}^{(1)},u_{2,i}^{(1)},s_{i}^{(2)}, u_{1,i}^{(2)}) \cdot \mathbbm{1}{ \bigl \{ {\ve H_{1}^{(2)}}\ve u_{1}^{(2)}=\ve{0} \bigr \}} 
\end{align}
where the equality $(h)$ follows from a similar argument as in (\ref{eq:u1_multilevel_level_1}). In particular, we can show that $p(\ve s^{(2)},\ve u_{1}^{(2)},\ve u_{1}^{(1)},\ve u_{2}^{(1)}) $ is equal to
\begin{align*}
\frac{ \mathbbm{1}\bigl\{ \ve{s}^{(2)} \oplus \ve{c}^{(2)} \in \tilde{\mathcal C}^{(2)} \bigr\}  \mathbbm{1}\bigl\{ \ve{u}_1^{(2)} \in \mathcal{C}_1^{(2)} \bigr\} \mathbbm{1}\bigl\{ \ve{u}_1^{(1)} \in \mathcal{C}_1^{(1)} \bigr\} \mathbbm{1}\bigl\{ \ve{u}_2^{(1)} \in \mathcal{C}_2^{(1)} \bigr\} }{ |\tilde{\mathcal C}^{(2)}|\,|\mathcal{C}_1^{(2)}|\,|\mathcal{C}_1^{(1)}|\,|\mathcal{C}_2^{(1)}|} 
\end{align*}
where $\ve s^{(2)}\oplus \ve c^{(2)}$, $\ve u_1^{(1)}$ and $\ve u_2^{(1)}$ are guaranteed to be codewords as they are decoded from the previous steps. Furthermore, the corresponding initial LLR value is given by
$\LLR_2^{(2)} =\log \frac{p(y_i| u_{1,i}^{(1)}, u_{2,i}^{(1)}, s_{i}^{(2)}, u_{1,i}^{(2)}=0)}{p(y_i|  u_{1,i}^{(1)}, u_{2,i}^{(1)}, s_{i}^{(2)}, u_{1,i}^{(2)}=1)}$
with the explicit expression given by (\ref{eqn:LLR22}) in  Appendix \ref{appen:derivations_LLR_multilevel}.

The consecutive decoding procedure should be clear from the foregoing two steps. In general, for any $\ell\geq 2$, we decode the bit string $\ve s^{(\ell)}$ in the sum $\ve s$, and one of the codewords, say $\ve u_{1}^{(\ell)}$.  To simplify the notation we use $\ve u_{k}^{(1:\ell)}$ to denote the set of vectors
$\ve u_{k}^{(1:\ell)}:=(\ve u_{k}^{(1)},\ve u_{k}^{(2)},\ldots, \ve u_{k}^{(\ell)}) \text{ for  } k=1,2.$
Based on the decoded bit strings $\ve u_{1}^{(1:\ell)}$ and $\ve u_{2}^{(1:\ell)}$, we have
\begin{align}
{\hat{s}}_{i}^{(\ell)} &=\argmax_{s_{i}^{(\ell)} \in \left\{0,1\right\}}p(s_{i}^{(\ell)}|\ve y,\ve u_{1}^{(1:\ell-1)},\ve u_{2}^{(1:\ell-1)}) \nonumber \\
&= \argmax_{s_{i}^{(\ell)} \in \left\{0,1\right\}} \sum_{\sim s_{i}^{(\ell)}}p(\ve y|\ve s^{(\ell)}, \ve u_{1}^{(1:\ell-1)},\ve u_{2}^{(1:\ell-1)})  \cdot p(\ve s^{(\ell)} |\ve u_{1}^{(1:\ell-1)},\ve u_{2}^{(1:\ell-1)}) \cdot p(\ve u_{1}^{(1:\ell-1)},\ve u_{2}^{(1:\ell-1)})  \nonumber \\
&\overset{(i)}{=} \argmax_{s_{i}^{(\ell)} \in \left\{0,1\right\}} \sum_{\sim s_{i}^{(\ell)}} \prod_{i=1}^{n} p(y_{i}| s_{i}^{(\ell)}, u_{1,i}^{(1:\ell-1)}, u_{2,i}^{(1:\ell-1)})  \cdot \mathbbm{1} \bigl\{ {\tilde{\ve H}^{(\ell)}(\ve s^{(\ell)} \oplus_2 \ve c^{(\ell)})=\ve 0} \bigr\}
\end{align}
To see why the last step holds, first observe that $\ve u_{1}^{(1: \ell -1)}, \ve u_{2}^{(1: \ell -1)}$ are guaranteed to be codewords as they are decoded from the previous steps. Furthermore, with decoded $\ve u_{1}^{(1:\ell-1)}, \ve u_{2}^{(1:\ell-1)}$, we define $\ve c^{(\ell)}$ as
\begin{align*}
c_{i}^{(\ell)}=\begin{cases}
0, &\text{if } \sum_{j=1}^{\ell-1}2^{j-1}( u_{1,i}^{(j)}+ u_{2,i}^{(j)})<2^{\ell-1}\\
1, &\text{if }  \sum_{j=1}^{\ell-1}2^{j-1}( u_{1,i}^{(j)}+ u_{2,i}^{(j)})\geq 2^{\ell-1}\
\end{cases}
\end{align*}
and observe that
$\ve s^{(\ell)}=\ve u_{1}^{(\ell)}\oplus_2\ve u_{2}^{(\ell)}\oplus_2 \ve c^{(\ell)}$
according to Lemma~\ref{lemma:relationship}. Using a similar argument as in (\ref{eq:ps_multiplevel_level_2}), we have
\begin{align}
p(\ve s^{(\ell)} |\ve u_{1}^{(1:\ell-1)},\ve u_{2}^{(1:\ell-1)})=\frac{1}{|\tilde{\mathcal C}^{(\ell)} |} \mathbbm{1} \bigl\{ {\tilde{\ve H}^{(\ell)}(\ve s^{(\ell)} \oplus \ve c^{(\ell)})=\ve 0} \bigr\}
\end{align}
where we have used the same arguments as in $(d)$-$(g)$. By substituting back to $(i)$ we have a complete explanation on decoding the bit string $\ve s^{(\ell)}$ with the initial LLR value for sum-product algorithm given by
$\LLR_1^{(\ell)}= \log \frac{p(y_{i}| u_{1,i}^{(1:\ell-1)}, u_{2,i}^{(1:\ell-1)},s_{i}^{(\ell)} \oplus c_{i}^{(\ell)}=0)}{p(y_{i}| u_{1,i}^{(1:\ell-1)}, u_{2,i}^{(1:\ell-1)},s_{i}^{(\ell)} \oplus c_{i}^{(\ell)}=1)}.$
Similarly we decode, say  $\ve u_{1}^{(\ell)}$, as follows
\begin{align} 
{\hat{u}}_{1,i}^{(\ell)}&=\argmax_{u_{1,i}^{(\ell)} \in \left\{0,1\right\}}p(u_{1,i}^{(\ell)}|\ve y,\ve u_{1}^{(1:\ell-1)},\ve u_{2}^{(1:\ell-1)},\ve s^{(\ell)} ) \nonumber \\
&= \argmax_{u_{1,i}^{(\ell)} \in \left\{0,1\right\}} \sum_{\sim u_{1,i}^{(\ell)}}p(\ve y|\ve s^{(\ell)}, \ve u_{1}^{(1:\ell)},\ve u_{2}^{(1:\ell-1)}) \cdot p(\ve s^{(\ell)} ,\ve u_{1}^{(1:\ell)},\ve u_{2}^{(1:\ell-1)}) \nonumber \\
&\overset{(j)}{=}\argmax_{u_{1,i}^{(\ell)} \in \left\{0,1\right\}} \sum_{\sim u_{1,i}^{(\ell)}} \prod_{i=1}^{n} p(y_{i}| s_{i}^{(\ell)}, u_{1,i}^{(1:\ell)}, u_{2,i}^{(1:\ell-1)}) \cdot \mathbbm{1} \bigl\{ {{\ve H_{1}^{(\ell)}}\ve u_{1}^{(\ell)}=\ve{0}} \bigr\}
\end{align}
where $(j)$ follows from similar argument as in (\ref{eq:ps_multiplevel_level_2}) together with the fact that $\ve s^{(\ell)}\oplus \ve c^{(\ell)}$, $\ve u_1^{(1: \ell-1)}$ and $\ve u_2^{(1:\ell -1)}$ are guaranteed to be codewords as they are decoded from the previous steps. 

The explicit expressions of $LLR_1^{(\ell)}, LLR_2^{(\ell)}$ is given in the following proposition.
\begin{proposition}[LLR for CFMA with multilevel binary codes]
The LLR values for the sum-product algorithm in the $\ell$-th level are given as
\begin{align}
\LLR_1^{(\ell)}&=\begin{cases}
		\LLR_{*1}^{(\ell)} & \text{for $c_i^{(\ell)}=0$} \\
		-\LLR_{*1}^{(\ell)} & \text{for $c_i^{(\ell)}=1$}
	\end{cases}
	\label{eq:LLR_multilevel}
\end{align}
where 
$\LLR_{*1}^{(\ell)} = \log \frac{p(y_{i}| u_{1,i}^{(1:\ell-1)}, u_{2,i}^{(1:\ell-1)},s_{i}^{(\ell)}=0)}{p(y_{i}| u_{1,i}^{(1:\ell-1)}, u_{2,i}^{(1:\ell-1)},s_{i}^{(\ell)}=1)},$
and the conditional probability density function is
\begin{align}
&p(y_{i}|s_{i}^{(\ell)}, u_{1,i}^{(1:\ell-1)}, u_{2,i}^{(1:\ell-1)})=\sum_{s_i^{(\ell +1:L)},u_{1,i}^{(\ell:L)}} \frac{1}{\sqrt{2 \pi}} \exp{\left(-\frac{1}{2}(y_i-\varphi_1(u_{1,i})-\varphi_2(u_{1,i} \oplus_{2^L} s_i))^2\right)}.
\end{align}
Furthermore,
$\LLR_2^{(\ell)} =\log \frac{p(y_i| u_{1,i}^{(1: \ell -1)}, u_{2,i}^{(1: \ell -1)}, s_{i}^{(\ell)}, u_{1,i}^{(\ell)}=0)}{p(y_i|  u_{1,i}^{(1: \ell -1)}, u_{2,i}^{(1: \ell -1)}, s_{i}^{(\ell)}, u_{1,i}^{(\ell)}=1)},$
and the conditional probability density function is
\begin{align}
&p(y_{i}|s_{i}^{(\ell)}, u_{1,i}^{(1:\ell)}, u_{2,i}^{(1:\ell-1)})=\sum_{s_i^{(\ell +1:L)},u_{1,i}^{(\ell+1:L)}} \frac{1}{\sqrt{2 \pi}} \exp{\left(-\frac{1}{2}(y_i-\varphi_1(u_{1,i})-\varphi_2(u_{1,i} \oplus_{2^L} s_i))^2\right)}.
\end{align}
\end{proposition}
Partial derivation is given in Appendix \ref{appen:derivations_LLR_multilevel}.

\begin{algorithm}[h!]
\caption{CFMA: Decoding Algorithm with multilevel codes}
\begin{algorithmic}[1]
\State $\LLR_1^{(1)}, \LLR_2^{(1)}$ \Comment{Initialize LLR values}  
\For{$\ell=1$ to $L$} 
	\State $\hat{\ve{s}}^{(\ell)} = \mathrm{SPA}(\tilde{\ve{H}}^{(\ell)}, \LLR_1^{(\ell)}$)\Comment{Decode $\ve s^{(\ell)}$}
	\State $\hat{\ve{u}}_1^{(\ell)} = \mathrm{SPA}(\ve H_1^{(\ell)}, \LLR_2^{(\ell)}$)\Comment{Decode $\ve{u}_1^{(\ell)}$}
	\State $\hat{\ve{u}}_2^{(\ell)} = \hat{\ve{s}}^{(\ell)} \oplus \hat{\ve{u}}_1^{(\ell)}$ \Comment{Decode $\ve{u}_2^{(\ell)}$}
    \State \text{Calculate }   $\LLR_1^{(\ell+1)}, \LLR_2^{(\ell+1)}$ \text{using (\ref{eq:LLR_multilevel})}
\EndFor
\State $\hat{\ve{s}} = \sum_{\ell=1}^L 2^{\ell-1} \hat{\ve{s}}^{(\ell)}$ \Comment{Recover the sum codeword $\ve{s}$}
\State $\hat{\ve{u}}_1 = \sum_{\ell=1}^L 2^{\ell-1} \hat{\ve{u}}_1^{(\ell)}$ \Comment{Recover the codeword $\ve{u}_1$}
\State $\hat{\ve{u}}_2 = \hat{\ve{s}} \oplus \hat{\ve{u}}_1$ \Comment{Recover the codeword $\ve{u}_2$}
		
\end{algorithmic}
\label{alg:cfma_multilevel}
\end{algorithm}

\begin{remark}
We note that in general, the users can use multilevel codes with different number of levels. Assume  user $1$ and user $2$ have $L_1$ and $L_2$ layers, respectively, where w.l.o.g $L_1\ge L_2$. The first $L_2$ levels follow the same procedure as the case with $L_1=L_2$. As a result, user $2$ is completely decoded and $L_2 -L_1$ levels of user $1$ remains to be decoded. Next we proceed by subtracting user $2$ and we are left with only a single-user channel.

\end{remark}

%
%

\section{Extensions}
In this section, we briefly discuss possible extensions of the CFMA framework to complex channels and the multi-user case. 

\subsection{CFMA for complex channel} \label{sec:CFMA complex}
The complex channel model is given by
	$Y = h_1 x_1 + h_2 x_2 + Z$,
where $h_1,h_2 \in \mathbb{C}$ denote the complex-valued channel gain and $
Z\sim\mathcal{CN}(0,1)$ is the circularly symmetric complex additive white Gaussian noise.

Conceptually, the extension can be done simply by mapping the odd and even code symbols to the complex and real channels, respectively. Formally, we have
\begin{align} \label{eqn:mapmulti}
(x_{1,i},x_{2,i}) &= (\varphi_1(u_{1,{2i-1}},u_{1,{2i}}),\varphi_2(u_{2,{2i-1}},u_{2,{2i}}))
\end{align}
where,
\begin{IEEEeqnarray*}{rCl}   
	\varphi_1(u_{1,{2i-1}},u_{2,{2i}}) &=& \sqrt{\frac{3P}{2(2^{2L}-1)}} \sum_{\ell=1}^L 2^{\ell-1} \times \left( \jmath (2 u_{1,{2i-1}}^{(\ell)}-1) + (2 u_{2,{2i}}^{(\ell)}-1) \right) \\
	\varphi_2(u_{2,{2i-1}},u_{2,{2i}}) &=& e^{j\theta} \cdot \varphi_1(u_{2,{2i-1}},u_{2,{2i}}).
\end{IEEEeqnarray*}

One notable feature in the above mapping is the parameter $\theta$ that relatively rotates the constellation between the two users. When we do not have channel state information at the transmitter (CSIT), the parameter $\theta$ is simply zero. In situations where we have CSIT, the parameter can further improve the performance. 
The parameter $\theta$ gives another degree of freedom to change the modulation (or the effective channel gain) between the users when we use QAM modulation. In some cases where the channel coefficients are unfavorable for compute-forward, the parameter $\theta$ can be used to tune the effective channel gains. By changing the parameter $\theta$ we can achieve more points on the dominant face of the MAC capacity. 
The decoding procedure is similar to that of the real-valued channel model developed  in previous sections \ref{sec:CFMA basic} and \ref{sec:CFMA extensions}, with the only exception that all LLR values will be evaluated with respect to the complex channel
 \begin{align*}
 p(y_i|u_{1,2i-1},u_{1,2i},u_{2,2i-1},u_{2,2i}) = \frac{1}{\pi} \exp\left(-\bigl| y_i - \varphi_1(u_{1,2i-1},u_{1,2i}) - \varphi_2(u_{2,2i-1},u_{2,2i}) \bigr|^2\right) \nonumber.
 \end{align*} 
By analogy, each decoding step also has a complexity of the order of the point-to-point communication system. Simulation results for complex-valued channel models are provided in Section \ref{sec:simulation_4QAM} and \ref{sec:simulation_16QAM}. 

\subsection{Multi-user MAC}
In this section, we briefly discuss how to extend the proposed CFMA strategy to the case of more than two users. We consider the multi-user Gaussian MAC, with input alphabets $\Xc_1$, $\Xc_2$, $\dots$, $\Xc_K$ and output alphabet $\Yc$. For input symbols $x_1$, $x_2$, $\dots$, $x_K$ the output symbol is given by $Y= \sum_{k=1}^K h_k x_k + Z,$
where $h_1,h_2, \dots, h_K \in \mathbb{R}$ denote the constant channel and $
Z\sim\mathcal{N}(0,1)$ is the additive white Gaussian noise. Any linear combination of the codewords has the following form $\bigoplus_{k=1}^K a_k \ve u_k.$ where $\ve u_k$ denotes the codeword of user $k$. For simplicity we denote 
	$\ve e^m=\bigoplus_{k=1}^m \ve u_k$ for $m \in \left\{2,3, \dots, K \right\}$.
In principle, any $K$ linearly independent linear combinations allow us to decode messages from all $K$ users. Here describe the decoding procedure with a specific choice of the linear combinations.   Particularly, in the  first stage  we perform bit-wise  estimation of $\ve e^K$ (the sum of all codewords), in the second stage we decode $\ve e^{K-1}$ (the sum of $K-1$ codewords), and so on. In the last stage we decode $\ve u_1$ i.e.
\begin{itemize}
\item{\makebox[2cm]{Decode $\ve{e}^K$: \hfill} ${\hat{e}^K}_{i}=\argmax\limits_{e^K_{i}}p(e^K_i|\ve{y})$} \\ 
\vspace{-2em} \vdots \\ 
\item{\makebox[2cm]{Decode $\ve{e}^{2}$: \hfill} ${\hat{e}}^{2}_{i}=\argmax\limits_{e^2_{i}}p(e^{2}_{i}|\ve{y},\ve{e}^K,\ve{e}^{K-1}, \dots, \ve{e}^3)$} \\ \vspace{-1em}
\item{\makebox[2cm]{Decode $\ve{u}_1$: \hfill} ${\hat{u}}_{1,i}=\argmax\limits_{u_{1,i}}p(u_{1,i}|\ve{y},\ve{e}^K,\ve{e}^{K-1}, \dots,  \ve{e}^2)$} 
\end{itemize}
For each decoding stage, the bit-wise MAP rule (in particular, the LLR values) can be derived in a similar way as in Section \ref{sec:4b}, and we omit the details here.


\section{Numerical simulations}  \label{sec:simulation}
In this section, we provide numerical simulations of our proposed CFMA codes and compare with the theoretical rate regions. For the theoretical rate regions, we evaluate theorem \ref{thm:cfma-nlc} with different discrete inputs.
Throughout our simulations we use the sum-product algorithm with $25$ iterations. We assume that a rate pair $(R_1,R_2)$ is achieved for a given power and channel gains if the bit-error rate (BER) is below \num{e-5} over \num{500} independent trials. We compare with the theoretical power P in dB that achieves the target rate pair in Theorem~\ref{thm:cfma-nlc} evaluated with discrete inputs. For all the simulations we always use $(a_1,a_2)=(1,1)$ for the first decoding step and $(a_1,a_2)=(1,0)$ for the second decoding step. We do not try to optimize the size of blocklength in different simulations.

For ease of presentation, for a fixed input distribution, we will reference the corner points of $\Rc_{\sf MAC-UI}$ as points $\mathsf{A}$ and $\mathsf{B}$, and the the corner points of $\Rc_{\sf CFMA-UI}$ by $\mathsf{A}'$ and $\mathsf{B}'$ (c.f. Figure~\ref{fig:mac-nlc}).

\subsection{CFMA: binary codes with BPSK modulation} \label{sec:6a}
In this scenario we set the power of the users to be the same $P_1=P_2$ and the real-valued channel gain pair to be $(h_1,h_2)=(1,\sqrt{3})$. The target rate pair is $(R_1, R_2)=(0.9742, 0.9355)$ which corresponds to the point $\mathsf{B}'$ shown in Figure \ref{fig:sim-theoretical1}. The LDPC blocksize is \num{4376}. For this particular setup, point $\mathsf{B}$ and $\mathsf{B}'$ coincide at one of the corner points of rate region~\eqref{eq:mac-cap}. 
The theoretical rate regions and performance evaluation for this case is given in Figures~\ref{fig:sim-theoretical1} and~\ref{fig:sim-practical1}, respectively.
The base LDPC code we use to construct our CFMA code is rate $R=0.9355$, and under the point-to-point AWGN channel, the code itself has a \SI{1.03}{\decibel} gap from the Shannon limit when used in a point-to-point AWGN channel. On the other hand, our CFMA strategy has \SI{1.5}{\decibel} gap from the corresponding theoretical bound.

\begin{figure}[!ht] 
\begin{minipage}[b]{0.48\textwidth}
\begin{tikzpicture}[every path/.append style={thick},scale=0.9, every node/.style={scale=0.9}]
\pgfplotstableread{
	0.974245777593474	0
	0.974245777593474	0.935473341276744
	0.935473350101363	0.974245768768855
	0	0.999964471359122
    }{\mytable};
\begin{axis}[
	xmin=0.75, ymin=0.75,
    ymax=1.03,
	xlabel={$R_1$ [bpcu]}, ylabel={$R_2$ [bpcu]},
	xmajorgrids, ymajorgrids, 
    axis x discontinuity=parallel,
    axis y discontinuity=parallel,
    axis equal,
    disabledatascaling=true,
	axis on top=true,
	]
    \coordinate (origin) at (0,0);
	\coordinate (BB) at (0.97424576876885,0.93547335010136);
	\coordinate (B) at (0.97424577759347,0.93547334127674);
	\coordinate (A) at (0.90975464751109,0.99996447135912);
	\coordinate (AA) at (0.93547335010136,0.97424576876885);
    \fill[black!10] (origin)-|(B)--(A)-|cycle;
    \fill[black!20] (origin)-|(BB)-|(AA)-|cycle;
  	\addplot [thick,dashed,draw=none] table {\mytable};
    \draw[thick,dashed] (AA-|origin)--(AA)--(AA|-origin);
    \draw[thick,dashed] (BB-|origin)--(BB)--(BB|-origin);
    \draw[<-,shorten <=2pt] (BB) .. controls +(-.02,-.03) and +(.01,.03) .. ($(BB)+(-.04,-.09)$) node[below left,anchor=north] {(0.9742, 0.9355)};
	\node[draw,fill=white,shape=circle,inner sep=1.2pt,label={[label distance={.9ex}]60:\textsf{A}}] at (A) {};
	\node[draw,fill=white,shape=circle,inner sep=1.2pt,label={[label distance={.2ex}]80:\textsf{A'}}] at (AA) {};
	\node[draw,fill=white,shape=circle,inner sep=1.2pt,label={[label distance={.2ex}]30:\textsf{B,B'}}] at (BB) {};
\end{axis}
\end{tikzpicture}
\caption{Target rate pair $\mathsf{B}'$ for the BPSK modulation case. The point $\mathsf{B}'$ in this case coincides with the corner point $\mathsf{B}$ of $\Rc_{\sf MAC-UI}$. Note that axes are cropped.}
\label{fig:sim-theoretical1}
\end{minipage}
\hfill
\begin{minipage}[b]{0.48\textwidth}
\begin{tikzpicture}[thick,scale=0.9, every node/.style={scale=0.9}]
\begin{axis}[%
xmin=3,
xmax=10,
xlabel={P[dB]},
ymode=log,
ymin=1e-05,
ymax=10,
ylabel={BER},
title={},
legend style={at={(0.405,0.6777)}, anchor=south west, legend cell align=left, align=left, draw=white!15!black, style={font=\small},}
]
\addplot [color=red, line width=1.0pt,mark=asterisk]
  table[row sep=crcr]{%
1	0.13198491773309\\
2	0.104564899451554\\
3	0.0788747714808044\\
4	0.0565635283363803\\
5	0.0378642595978062\\
6	0.0230379341864717\\
7	0.0124195612431444\\
7.5	0.00831398537477148\\
8	0.00421389396709324\\
8.2	0.00246160877513711\\
8.4	0.000940127970749543\\
8.6	0.000251828153564899\\
9	9.140767824497258e-06\\
9.2	1.599634369287020e-06\\
9.4	0\\
};
\addlegendentry{Decoding $\ve{s}$}

\addplot [color=blue, line width=1.0pt, dashed, mark=+,mark options={solid}] table[row sep=crcr]{%
1	0.102931901279707\\
2	0.0889181901279708\\
3	0.0746946983546618\\
4	0.0608811700182815\\
5	0.0479680073126143\\
6	0.035412705667276\\
7	0.0244753199268739\\
7.5	0.0191864716636197\\
8	0.0133765996343693\\
8.2	0.0105900365630713\\
8.4	0.00631535648994516\\
8.6	0.00291133455210238\\
9	2.461151736745887e-04\\
9.2	4.113345521023766e-05\\
9.4	9.882632541133460e-06\\
};
\addlegendentry{Decoding $\ve{u}_1$ given $\ve{s}$}

\addplot [color=teal, line width=1.0pt, mark=x,mark options={solid}, dashdotted] table[row sep=crcr]{%
1	0.230964351005484\\
2	0.192214808043876\\
3	0.153252285191956\\
4	0.117385283363803\\
5	0.0858222120658135\\
6	0.058449268738574\\
7	0.0368948811700183\\
7.5	0.0275004570383912\\
8	0.0175904936014625\\
8.2	0.0130516453382084\\
8.4	0.0072554844606947\\
8.6	0.00316316270566728\\
9	2.552559414990859e-04\\
9.2	4.273308957952468e-05\\
9.4	9.882632541133460e-06\\
};
\addlegendentry{Decoding $\ve{u}_2 = \ve{s} \oplus \ve{u}_1$}


\addplot [color=black, line width=1pt] table[row sep=crcr]{%
-1	0.186156307129799\\
0	0.158839122486289\\
1	0.130143967093236\\
2	0.104620658135283\\
3	0.0788185557586837\\
4	0.0564488117001828\\
5	0.0381010054844607\\
6	0.021636197440585\\
6.4	0.0166142595978062\\
6.7	0.0115683272394881\\
7	0.00464808043875686\\
7.3	0.000679159049360146\\
7.5	0.000103519195612431\\
7.6	9.54113345521024e-06\\
7.7	1.03815356489945e-06\\
};

\addlegendentry{Point-to-point AWGN}

\addplot [color=black, line width=1.0pt] table[row sep=crcr]{%
6.57	0.00001\\
6.57	0.001\\
};

\addplot [color=black, line width=1.0pt, forget plot] table[row sep=crcr]{%
7.912	0.00001\\
7.912	0.001\\
};

\end{axis}
\draw[black] (1.8,2.2) node[anchor=north,text width=3cm, align=right]{Shannon \\ limit for \\ p2p};
\draw[black] (3.8,2) node[anchor=north,text width=3cm, align=right]{Theoretical \\ bound };

\end{tikzpicture}%
\caption{Bit error rate simulation results for each decoding step for BPSK modulation and target rate pair $(R_1, R_2)=(0.9742, 0.9355)$. For reference, we include the base code performance over the AWGN point-to-point channel (the black line).}\label{fig:sim-practical1}
\end{minipage}
\end{figure}
\vspace{-2em}
\subsection{CFMA: binary codes with 4-QAM modulation}
\label{sec:simulation_4QAM}

In this scenario we set the power of the users to be the same $P_1=P_2$ and the channel gains to be equal with unity $h_1=h_2=1$ 
(note that the input symbols are complex numbers since we use 4-QAM in this subsection). The target rate pair is $(R_1,R_2)=(1.885,1.871)$, and the LDPC blocksize is \num{4376}. 
The theoretical rate regions and performance evaluation for this case is given in Figures~\ref{fig:sim-theoretical2} and~\ref{fig:sim-practical2}, respectively.
The base code is the same as in section \ref{sec:6a}. In the simulation for this case, we observe that our CFMA strategy has \SI{1.7}{\decibel} gap from the corresponding theoretical power P. 

\begin{figure}[!ht]
\begin{minipage}[b]{0.48\textwidth}
\begin{tikzpicture}[every path/.append style={thick},scale=0.9, every node/.style={scale=0.9}]
\pgfplotstableread{
	1.99815617180102	0
	1.99815617180102	1.75719841696521
	1.75719841734026	1.99815617142598
	0               	1.99815617142598
    }{\mytable};
\begin{axis}[
	xmin=1.5, ymin=1.5,
	xlabel={$R_1$ [bpcu]}, ylabel={$R_2$ [bpcu]},
	xmajorgrids, ymajorgrids, 
    axis x discontinuity=parallel,
    axis y discontinuity=parallel,
    axis equal,
    disabledatascaling=true,
	axis on top=true,
	]
    \coordinate (origin) at (0,0);
	\coordinate (BB) at (1.88455574436148,1.87079884440475);
	\coordinate (B) at (1.99815617180102,1.75719841696521);
	\coordinate (A) at (1.75719841734026,1.99815617142598);
	\coordinate (AA) at (1.87079884440475,1.88455574436148);
    \fill[black!10] (origin)-|(B)--(A)-|cycle;
    \fill[black!20] (origin)-|(BB)-|(AA)-|cycle;
  	\addplot [thick,dashed,draw=none] table {\mytable};
    \draw[thick,dashed] (AA-|origin)--(AA)--(AA|-origin);
    \draw[thick,dashed] (BB-|origin)--(BB)--(BB|-origin);
    \draw[<-,shorten <=2pt] (BB) .. controls +(-.02,-.03) and +(.01,.03) .. ($(BB)+(-.04,-.09)$) node[below left,anchor=north] {(1.885,1.871)};
	\node[draw,fill=white,shape=circle,inner sep=1.2pt,label={[label distance={.2ex}]60:\textsf{A}}] at (A) {};
	\node[draw,fill=white,shape=circle,inner sep=1.2pt,label={[label distance={.2ex}]80:\textsf{A'}}] at (AA) {};
	\node[draw,fill=white,shape=circle,inner sep=1.2pt,label={[label distance={.2ex}]30:\textsf{B}}] at (B) {};
	\node[draw,fill=white,shape=circle,inner sep=1.2pt,label={[label distance={.2ex}]10:\textsf{B'}}] at (BB) {};
\end{axis}
\end{tikzpicture}
\caption{Target rate pair $\mathsf{B}'$ for the 4-QAM modulation case. The achievable rate pair $\mathsf{B}'$ of the CFMA strategy achieves a non-endpoint on the dominant face of $\Rc_{\sf MAC-UI}$. Note that axes are cropped.} 
\label{fig:sim-theoretical2}
\end{minipage}
\hfill
\begin{minipage}[b]{0.48\textwidth}
\begin{tikzpicture}[thick, scale=0.85, every node/.style={scale=0.85}]
\begin{axis}[%
xmin=5,
xmax=13,
xlabel={P[dB]},
ymode=log,
ymin=1e-05,
ymax=10,
ylabel={BER},
title={},
legend style={at={(0.405,0.6777)}, anchor=south west, legend cell align=left, align=left, draw=white!15!black, style={font=\small},}
]
\addplot [color=red, line width=1.0pt,mark=asterisk]
  table[row sep=crcr]{%
5	0.108777879341865\\
6	0.0853478062157221\\
7	0.065758226691042\\
8	0.0497481718464351\\
9	0.0355338208409506\\
10	0.0240744972577697\\
11	0.0150301645338208\\
11.5	0.0109031078610603\\
12	0.00655484460694698\\
12.1	0.00538025594149909\\
12.2	0.00382861060329068\\
12.3	0.00230438756855576\\
12.4	0.00109643510054845\\
12.5	0.000214351005484461\\
12.6	2.83363802559415e-05\\
12.7	3.74780174728101e-06\\
};
\addlegendentry{Decoding $\ve{s}$}

\addplot [color=blue, line width=1.0pt, dashed, mark=+,mark options={solid}] table[row sep=crcr]{%
5	0.204246800731261\\
6	0.173606489945155\\
7	0.14315676416819\\
8	0.114316727605119\\
9	0.0851316270566728\\
10	0.0603697440585009\\
11	0.0401732175502742\\
11.5	0.0311165447897623\\
12	0.0219986288848263\\
12.1	0.0190827239488117\\
12.2	0.0146165447897623\\
12.3	0.00943098720292504\\
12.4	0.00442138939670932\\
12.5	0.000932815356489945\\
12.6	0.000129798903107861\\
12.7	0.000009120365204028\\
};
\addlegendentry{Decoding $\ve{u}_1$ given $\ve{s}$}

\addplot [color=teal, line width=1.0pt, mark=x,mark options={solid}, dashdotted] table[row sep=crcr]{%
5	0.202229433272395\\
6	0.173425045703839\\
7	0.1439478976234\\
8	0.115021023765996\\
9	0.0857001828153565\\
10	0.0603446069469835\\
11	0.0401165447897623\\
11.5	0.0311412248628885\\
12	0.0221668190127971\\
12.1	0.0192554844606947\\
12.2	0.0147815356489945\\
12.3	0.00954159049360146\\
12.4	0.00454890310786106\\
12.5	0.000956124314442413\\
12.6	0.000129798903107861\\
12.7	0.000009120365204028\\
};
\addlegendentry{Decoding $\ve{u}_2 = \ve{s} \oplus \ve{u}_1$}

\addplot [color=black, line width=1pt] table[row sep=crcr]{%
-1	0.186156307129799\\
0	0.158839122486289\\
1	0.130143967093236\\
2	0.104620658135283\\
3	0.0788185557586837\\
4	0.0564488117001828\\
5	0.0381010054844607\\
6	0.021636197440585\\
6.4	0.0166142595978062\\
6.7	0.0115683272394881\\
7	0.00464808043875686\\
7.3	0.000679159049360146\\
7.5	0.000103519195612431\\
7.6	9.54113345521024e-06\\
7.7	1.03815356489945e-06\\
};
\addlegendentry{Point-to-point AWGN}

\addplot [color=black, line width=1.0pt] table[row sep=crcr]{%
6.57	0.00001\\
6.57	0.001\\
};

\addplot [color=black, line width=1.0pt, forget plot] table[row sep=crcr]{%
10.93	0.00001\\
10.93	0.001\\
};

\end{axis}
\draw[black] (-0.2,2.2) node[anchor=north,text width=3cm, align=right]{Shannon \\ limit for \\ p2p};
\draw[black] (3.5,2) node[anchor=north,text width=3cm, align=right]{Theoretical \\ bound };

\end{tikzpicture}%
\caption{Bit error rate simulation results for each decoding step for 4-QAM modulation and target rate pair $(R_1, R_2)=(1.885,1.871)$. For reference, we include the base code performance over the AWGN point-to-point channel.}\label{fig:sim-practical2}
\end{minipage}
\end{figure}

\subsection{CFMA: multilevel codes with 16-QAM modulation}
\label{sec:simulation_16QAM}

In this scenario we set the power of the user to be the same $P_1=P_2$ and the channel gains to be equal and unity $h_1=h_2=1$ (note that the input symbols are complex numbers since we use 16-QAM in this subsection). The theoretical rate regions and performance evaluation for this case is given in Figures~\ref{fig:sim-theoretical3} and~\ref{fig:sim-practical3}, respectively.
The target rate pair is $(R_1,R_2)=(3.864,3.555)$, and the LDPC blocksize is \num{1908}. We use a base LDPC code that has $\SI{1.41}{\decibel}$ difference from the theoretical power P for the point-to-point AWGN channel. We simulate the BER for decoding $\ve{u}_1$ and $\ve{u}_2$ as a function of P for this case. 

\begin{figure}[!ht]
\begin{minipage}[b]{0.45\textwidth}
\begin{tikzpicture}[every path/.append style={thick},scale=0.85, every node/.style={scale=0.85}]
\pgfplotstableread{
	3.99999963676631	0
	3.99999963676631	3.41914661819364
	3.41914661898605	3.9999996359739
	0	3.9999996359739
    }{\mytable};
\begin{axis}[
	xmin=3, ymin=3,
	xlabel={$R_1$ [bpcu]}, ylabel={$R_2$ [bpcu]},
	xmajorgrids, ymajorgrids, 
    axis x discontinuity=parallel,
    axis y discontinuity=parallel,
    axis equal,
    disabledatascaling=true,
    axis on top=true,
	]
	\coordinate (BB) at (3.86408938451388,3.55505687044607);
	\coordinate (B) at (3.99999963676631,3.41914661819364);
	\coordinate (A) at (3.41914661898605,3.9999996359739);
	\coordinate (AA) at (3.55505687044607,3.86408938451388);
    \fill[black!10] (origin)-|(B)--(A)-|cycle;
    \fill[black!20] (origin)-|(BB)-|(AA)-|cycle;
  	\addplot [thick,dashed,draw=none] table {\mytable};
    \draw[thick,dashed] (AA-|origin)--(AA)--(AA|-origin);
    \draw[thick,dashed] (BB-|origin)--(BB)--(BB|-origin);
    \draw[<-,shorten <=2pt] (BB) .. controls +(-.02,-.04) and +(.03,.03) .. ($(BB)+(-.06,-.09)$) node[below left,anchor=north east] {(3.864,3.555)};
	\node[draw,fill=white,shape=circle,inner sep=1.2pt,label={[label distance={.2ex}]60:\textsf{A}}] at (A) {};
	\node[draw,fill=white,shape=circle,inner sep=1.2pt,label={[label distance={.2ex}]80:\textsf{A'}}] at (AA) {};
	\node[draw,fill=white,shape=circle,inner sep=1.2pt,label={[label distance={.2ex}]30:\textsf{B}}] at (B) {};
	\node[draw,fill=white,shape=circle,inner sep=1.2pt,label={[label distance={.2ex}]10:\textsf{B'}}] at (BB) {};
\end{axis}
\end{tikzpicture}
\caption{Target rate pair $\mathsf{B}'$ for the 16-QAM modulation case. The achievable rate pair $\mathsf{B}'$ of the CFMA strategy achieves a non-endpoint on the dominant face of $\Rc_{\sf MAC-UI}$. Note that axes are cropped.} \label{fig:sim-theoretical3}
\end{minipage}
\hfill
\begin{minipage}[b]{0.5\textwidth}
\begin{tikzpicture}[thick,scale=0.85, every node/.style={scale=0.85}]
\definecolor{mycolor1}{rgb}{0.82000,0.24700,0.12100}%
\definecolor{mycolor2}{rgb}{0.00000,0.44700,0.74100}%
	\begin{semilogyaxis}[
	xmin=0,
	xmax=30,
	xlabel={P[dB]},
	ymode=log,
	ymin=1e-05,
	ymax=500,
	ylabel=BER,
	legend style={at={(0.405,0.6777)}, anchor=south west, legend cell align=left, align=left, draw=white!15!black, style={font=\small},}
]

\addplot [color=red, line width=1.0pt,mark=asterisk]
  table[row sep=crcr]{%
-5	1.3793605870021\\
0	1.24104192872117\\
5	0.985392033542977\\
10	0.717148846960168\\
15	0.54359748427673\\
20	0.266454926624738\\
25	0.0576750524109015\\
26	0.0338155136268344\\
27	0.013517819706499\\
27.5	0.00461635220125786\\
28	0.00069601677148847\\
28.2	0.000259958071278826\\
28.4	0.000009778284292401\\
};
\addlegendentry{Decoding $\ve{s}$}

\addplot [color=blue, line width=1.0pt, dashed, mark=+,mark options={solid}]
  table[row sep=crcr]{%
-5	1.3201572327044\\
0	1.26407756813417\\
5	1.14974423480084\\
10	1.03151362683438\\
15	0.818299790356394\\
20	0.387345911949686\\
25	0.0976100628930818\\
26	0.0521844863731656\\
27	0.0203815513626834\\
27.5	0.00550524109014675\\
28	0.000538784067085954\\
28.2	0.000255765199161426\\
28.4	0.000011720182423764\\
};
\addlegendentry{Decoding $\ve{u}_1$ given $\ve{s}$}

\addplot [color=teal, line width=1.0pt, mark=x,mark options={solid}, dashdotted]
  table[row sep=crcr]{%
-5	1.28701048218029\\
0	1.23776100628931\\
5	1.20669601677149\\
10	1.11910482180293\\
15	0.888125786163522\\
20	0.428201257861635\\
25	0.106073375262055\\
26	0.0614947589098533\\
27	0.0251090146750524\\
27.5	0.00713626834381551\\
28	0.000769392033542977\\
28.2	0.000289308176100629\\
28.4	0.000014982338902173\\
};
\addlegendentry{Decoding $\ve{u}_2 = \ve{s} \oplus \ve{u}_1$}

\addplot [color=black, line width=1pt]
  table[row sep=crcr]{%
-1	0.186462264150943\\
0	0.159093291404612\\
1	0.132106918238994\\
2	0.102861635220126\\
3	0.0787997903563941\\
4	0.0562997903563941\\
5	0.035083857442348\\
5.6	0.0238050314465409\\
6	0.0109329140461216\\
6.4	0.00169287211740042\\
6.6	4.675052410901468e-04\\
6.8	4.979035639412998e-05\\
6.9 	1.870020964360587e-05\\
7	3.144654088050315e-06\\
};
\addlegendentry{Point-to-point AWGN}

\addplot [color=black, line width=1pt, forget plot]
  table[row sep=crcr]{%
21.7	1e-05\\
21.7	0.001\\
};

\addplot [color=black, line width=1.0pt, forget plot]
  table[row sep=crcr]{%
5.53	0.00001\\
5.53	0.001\\
};
\end{semilogyaxis}
\draw[black] (-0.3,2.2) node[anchor=north,text width=3cm, align=right]{Shannon \\ limit for \\ p2p};
\draw[black] (3.4,2) node[anchor=north,text width=3cm, align=right]{Theoretical \\ bound };
\end{tikzpicture}
\caption{Bit error rate simulation results for each decoding step for 16-QAM modulation and target rate pair $(R_1, R_2)=(3.864,3.555)$. For reference, we include the base code performance over the AWGN point-to-point channel.}\label{fig:sim-practical3}
\end{minipage}
\end{figure}

\subsection{CFMA: Gaussian interference channel}
As discussed in \ref{sec:intro}, CFMA can also be employed in a Gaussian interference channel without further modification. In this section, we illustrate this implementation with numerical results. Consider the symmetric 2-user Gaussian interference channel (IC) which is given by
\begin{align}
Y_1&=x_1+hx_2+Z_1, \quad
Y_2=hx_1+x_2+Z_2,
\end{align} 
where $h$ is the cross channel gain of both users, $Z_k\sim\mathcal{N}(0,1)$ is the additive white Gaussian noise at receiver $k=1,2$. In this scenario we set $h=\sqrt{3}$  and assign the same power for both users $P_1=P_2$. For the interference channel, our main focus is the case when both receivers decode interference, i.e., each decoder treats the channel as a multiple-access channel. The achievable rate region for this case is the intersection of two Gaussian MAC rate regions: one composed of two transmitters and receiver 1; and the other one composed of two transmitters and receiver 2. Denote the two MAC regions by $\Rc^{(1)}_{\sf MAC-UI}$ and $\Rc^{(2)}_{\sf MAC-UI}$, where $\Rc^{(1)}_{\sf MAC-UI}$ and $\Rc^{(2)}_{\sf MAC-UI}$ are the achievable rate regions in~\eqref{eq:mac-cap} with uniform BPSK inputs for receiver 1 and 2, respectively. We also denote the CFMA rate region in~\eqref{eq:cfma-nlc} with uniform BPSK inputs of receiver 1 and receiver 2 by $\Rc^{(1)}_{\sf CFMA-UI}$ and $\Rc^{(2)}_{\sf CFMA-UI}$, respectively. 

In the following we further proceed with the help of Figure~\ref{fig:GIC_reg}. In Figure~\ref{fig:GIC_reg}, the corner points of $R^{(k)}_{\sf MAC-UI}$, $k=1,2,$ are denoted by points $\mathsf{A}_k$ and $\mathsf{B}_k$ and the corner points of $\Rc^{(k)}_{\sf CFMA-UI}$, $k=1,2,$ are denoted by $\mathsf{A}_k'$ and $\mathsf{B}_k'$. The target rate pair we wish to achieve is the rate pair corresponding to points $\mathsf{B}_1'$ or $\mathsf{B}_2'$ which coincide. We note that this rate pair is not achievable under conventional successive cancellation decoders. 
For the simulations we use an LDPC code with blocksize 4376. The performance evaluation for this case is given in Figure~\ref{fig:GIC-simulation}.
In the numerical simulation we observe that for this Gaussian IC example, we have a \SI{1.3}{\decibel} gap from the corresponding theoretical bound.

\begin{figure}
\begin{minipage}[b]{0.48\textwidth}
\begin{tikzpicture}[every path/.append style={thick},scale=0.9, every node/.style={scale=0.9}]
\pgfplotstableread{
	0.974245777593474	0
	0.974245777593474	0.935473341276744
	0.935473350101363	0.974245768768855
	0	0.999964471359122
    }{\mytable};
\begin{axis}[
	xmin=0.75, ymin=0.75,
    ymax=1.03,
	xlabel={$R_1$ [bpcu]}, ylabel={$R_2$ [bpcu]},
	xmajorgrids, ymajorgrids, 
    axis x discontinuity=parallel,
    axis y discontinuity=parallel,
    axis equal,
    disabledatascaling=true,
	axis on top=true,
	]
    \coordinate (origin) at (0,0);
	\coordinate (BB) at (0.97424576876885,0.93547335010136);
	\coordinate (B) at (0.97424577759347,0.93547334127674);
	\coordinate (A) at (0.90975464751109,0.99996447135912);
	\coordinate (AA) at (0.93547335010136,0.97424576876885);
	\coordinate (AA2) at (0.93547335010136,0.97424576876885);
	\coordinate (A2) at (0.93547334127674,0.97424577759347);
	\coordinate (B2) at (0.99996447135912,0.90975464751109);
	\coordinate (BB2) at (0.97424576876885,0.93547335010136);	
    \fill[black!10] (origin)-|(AA)--(A)-|cycle;
    \fill[black!10] (origin)-|(B2)--(BB2)-|cycle;
    \fill[black!10] (origin)-|(BB)--(AA)-|cycle;
    \fill[black!20] (origin)-|(BB)-|(AA)-|cycle;
  	\addplot [thick,dashed,draw=none] table {\mytable};
    \draw[thick,dashed] (AA-|origin)--(AA)--(AA|-origin);
    \draw[thick,dashed] (BB-|origin)--(BB)--(BB|-origin);
    \draw[<-,shorten <=2pt] (BB) .. controls +(-.02,-.03) and +(.01,.03) .. ($(BB)+(-.04,-.09)$) node[below left,anchor=north] {(0.9742, 0.9355)};
	\node[draw,fill=white,shape=circle,inner sep=1.2pt,label={[label distance={.9ex}]45:$\mathsf{A}_1$}] at (A) {};
	\node[draw,fill=white,shape=circle,inner sep=1.2pt,label={[label distance={.2ex}]45:$\mathsf{A}_1'$, $\mathsf{A}_2$, $\mathsf{A}_2'$}] at (AA) {};
	\node[draw,fill=white,shape=circle,inner sep=1.2pt,label={[label distance={.2ex}]45:$\mathsf{B}_1$, $\mathsf{B}_1'$, $\mathsf{B}_2'$}] at (BB) {};
	\node[draw,fill=white,shape=circle,inner sep=1.2pt,label={[label distance={.2ex}]45:$\mathsf{B}_2$}] at (B2) {};
\end{axis}
\end{tikzpicture}
\caption{Achievable rate regions for the symmetric Gaussian interference channel. The pentagon with corner points $\mathsf{A}_k$ and $\mathsf{B}_k$ corresponds to the rate region~\eqref{eq:mac-cap} of receiver $k=1,2$ with uniform BPSK inputs. The union of the rectangular rate regions with corner points $\mathsf{A}_k'$ and $\mathsf{B}_k'$ correspond to the theoretical CFMA rate region~\eqref{eq:cfma-nlc} of receiver $k=1,2,$ with uniform BPSK inputs. The target rate pair for the Gaussian interference channel is $\mathsf{B}_1'$ (overlapped with $\mathsf{B}_2'$). Note that axes are cropped.}
\label{fig:GIC_reg}
\end{minipage}
\hfill
\begin{minipage}[b]{0.45\textwidth}
\begin{tikzpicture}[thick,scale=0.9, every node/.style={scale=0.9}]
\begin{axis}[%
xmin=3,
xmax=10,
xlabel={P[dB]},
ymode=log,
ymin=1e-05,
ymax=10,
ylabel={BER},
title={},
legend style={at={(0.405,0.6777)}, anchor=south west, legend cell align=left, align=left, draw=white!15!black, style={font=\small},}
]
\addplot [color=red, line width=1.0pt,mark=asterisk]
  table[row sep=crcr]{%
1	0.132286106032907\\
2	0.10420155393053\\
3	0.0786700182815357\\
4	0.0565105118829982\\
5	0.0377705667276051\\
6	0.023050731261426\\
7	0.0124140767824497\\
7.5	0.00851736745886655\\
8	0.00439031078610603\\
8.2	0.00241042047531993\\
8.4	0.000996800731261426\\
8.6	0.000283363802559415\\
9	1.55393053016453e-05\\
9.2	2.74223034734918e-06\\
9.5	0\\
};
\addlegendentry{Decoding $\ve{s}$}

\addplot [color=blue, line width=1.0pt, dashed, mark=+,mark options={solid}] table[row sep=crcr]{%
1	0.101143510054845\\
2	0.0873884826325411\\
3	0.0735863802559415\\
4	0.0599830895795247\\
5	0.0471384826325411\\
6	0.0349721206581353\\
7	0.0242001828153565\\
7.5	0.0187856489945155\\
8	0.0128967093235832\\
8.2	0.00996160877513711\\
8.4	0.00571252285191956\\
8.6	0.00230712979890311\\
9	0.000140310786106033\\
9.2	9.59780621572212e-06\\
9.5	0\\
};
\addlegendentry{Decoding $\ve{u}_1$ given $\ve{s}$}

\addplot [color=teal, line width=1.0pt, mark=x,mark options={solid}, dashdotted] table[row sep=crcr]{%
1	0.229562614259598\\
2	0.190347806215722\\
3	0.151962979890311\\
4	0.116444241316271\\
5	0.0848985374771481\\
6	0.0580228519195612\\
7	0.036613802559415\\
7.5	0.0273030164533821\\
8	0.0172870201096892\\
8.2	0.012372029250457\\
8.4	0.00670932358318099\\
8.6	0.00259049360146252\\
9	0.000155850091407678\\
9.2	1.23400365630713e-05\\
9.5	0\\
};

\addlegendentry{Decoding $\ve{u}_2 = \ve{s} \oplus \ve{u}_1$}

\addplot [color=black, line width=1pt] table[row sep=crcr]{%
-1	0.186156307129799\\
0	0.158839122486289\\
1	0.130143967093236\\
2	0.104620658135283\\
3	0.0788185557586837\\
4	0.0564488117001828\\
5	0.0381010054844607\\
6	0.021636197440585\\
6.4	0.0166142595978062\\
6.7	0.0115683272394881\\
7	0.00464808043875686\\
7.3	0.000679159049360146\\
7.5	0.000103519195612431\\
7.6	9.54113345521024e-06\\
7.7	1.03815356489945e-06\\
};
\addlegendentry{point-to-point AWGN}

\addplot [color=black, line width=1.0pt] table[row sep=crcr]{%
6.57	0.00001\\
6.57	0.001\\
};

\addplot [color=black, line width=1.0pt, forget plot] table[row sep=crcr]{%
7.912	0.00001\\
7.912	0.001\\
};

\end{axis}
\draw[black] (1.8,2.2) node[anchor=north,text width=3cm, align=right]{Shannon \\ limit for \\ p2p};
\draw[black] (5,2) node[anchor=north,text width=3cm, align=right]{Theoretical \\ bound };

\end{tikzpicture}%
\caption{Numerical evaluation for the Gaussian IC which depicts the bit error rate for each decoding stage at both receivers. For reference, we include the base code performance over the AWGN point-to-point channel.} \label{fig:GIC-simulation}
\end{minipage}
\end{figure}

\section{Conclusion}
In this paper, we presented a practical CFMA coding strategy with low complexity sequential decoders. We have shown that the CFMA strategy achieves (non-corner) points on the dominant face MAC capacity region without time-sharing or rate-splitting. This property leads to more flexible rate (resource) allocations for multi-user networks and rate improvement for applications such as the interference channel. Several case studies have been presented with off-the-self point-to-point binary LDPC codes that show the potential of our strategy. This property itself is desirable in many cases where backward compatibility is an issue. As future work, it would be interesting to see how the performance of the CFMA strategy can be improved by further optimization of the codes.   

%

\appendix
In this appendix we provide detailed derivations and expressions of conditional probability density functions and log-likelihood ratio values (LLR).

\subsection{Derivations of (\ref{eq:LLR_basic_CFMA}).}
\label{appen:derivations_LLR_basic}
The derivations of the LLR values in (\ref{eq:LLR_basic_CFMA}) are given in this section. The channel is AWGN in which
\begin{align}
p(y_i|u_{1,i},u_{2,i})= \frac{1}{\sqrt{2 \pi}} \exp \left(-\frac{1}{2} (y_i - x_{1,i}- x_{2,i})^2 \right).
\end{align}

Thus,
\begin{subequations}
\begin{IEEEeqnarray*}{rCl}
	\LLR_1
	&=& \log \frac{p(y_i|s_i=0)}{p(y_i|s_i=1)} = \log \frac{p(y_i|u_{1,i}=0,u_{2,i}=0)+p(y_i|u_{1,i}=1,u_{2,i}=1)}{p(y_i|u_{1,i}=0,u_{2,i}=1)+p(y_i|u_{1,i}=1,u_{2,i}=0)} \\
	&=& \log \frac{e^{-\frac{1}{2}(y_i+2\sqrt{P})^2}+e^{-\frac{1}{2}(y_i-2\sqrt{P})^2}}{2e^{-\frac{1}{2}y_i^2}} = \log \frac{e^{y_i2\sqrt{P}-2P}+e^{-y_i2\sqrt{P}-2P}}{2} \\
	 &=& \log\cosh\bigl(y_i2\sqrt{P}\bigr) - 2P \IEEEyesnumber \\
	\LLR_2
	 &=& \log \frac{p(y_i|u_{1,i}=0,s_i)}{p(y_i|u_{1,i}=1,s_i)} = \log \frac{p(y_i|u_{1,i}=0,u_{2,i}=s_i)}{p(y_i|u_{1,i}=1,u_{2,i}=1\oplus s_i)} \\
	 &=&
	 	\begin{cases}
			\log \frac{e^{-\frac{1}{2}(y_i-2\sqrt{P})^2}}{e^{-\frac{1}{2}(y_i+2\sqrt{P})^2}} & \text{for $s_i=0$} \\
            \log \frac{e^{-\frac{1}{2}y_i^2}}{e^{-\frac{1}{2}y_i^2}}& \text{for $s_i=1$}
		\end{cases} \quad =
	 	\begin{cases}
			4 y_i\sqrt{P} & \text{for $s_i=0$} \\
			0 & \text{for $s_i=1$}.
		\end{cases}   \IEEEyesnumber
\end{IEEEeqnarray*}
\end{subequations}

\subsection{Derivations on Section \ref{sec:CFMA extensions}.}
\label{appen:derivations_LLR_multilevel}
The conditional probability density function necessary for deriving $\LLR_1^{(1)}$ is 
\begin{align} 
&p(y_i|s_i^{(1)})=\sum_{s_i^{(2:L)}} p(y_i|s_i) = \sum_{s_i^{(2:L)},u_{1,i}^{(1:L)}} p(y_i|u_{1,i},u_{2,i}=u_{1,i} \oplus_{2^L} s_i) \\
&= \sum_{s_i^{(2:L)},u_{1,i}^{(1:L)}} \frac{1}{\sqrt{2 \pi}} \exp{\left(-\frac{1}{2}(y_i-\varphi_1(u_{1,i})-\varphi_2(u_{1,i} \oplus_{2^L} s_i))^2\right)}. \nonumber
\end{align}
The conditional probability density function that describes $\LLR_2^{(1)}$ is
\begin{align}
\label{eqn:LLR21}
&p(y_i|s_i^{(1)},u_{1,i}^{(1)})= \sum_{s_i^{(2:L)},u_{1,i}^{(2:L)}} p(y_i|u_{1,i},u_{2,i}=u_{1,i} \oplus_{2^L} s_i) \\
&=\sum_{s_i^{(2:L)},u_{1,i}^{(2:L)}} \frac{1}{\sqrt{2 \pi}} \exp{\left(-\frac{1}{2}(y_i-\varphi_1(u_{1,i})-\varphi_2(u_{1,i} \oplus_{2^L} s_i))^2\right)}. \nonumber
\end{align}
For the derivations of $\LLR_1^{(2)}$, we can split it as follows 
\begin{align}
\LLR_1^{(2)}&=\log \frac{p(y_i| u_{1,i}^{(1)}, u_{2,i}^{(1)}, s_{i}^{(2)} \oplus c_i^{(2)}=0)}{p(y_i|  u_{1,i}^{(1)}, u_{2,i}^{(1)}, s_{i}^{(2)} \oplus c_i^{(2)}=1)} =\begin{cases}
		\LLR_{*1}^{(2)} & \text{for $c_i^{(2)}=0$} \\
		-\LLR_{*1}^{(2)} & \text{for $c_i^{(2)}=1$}
	\end{cases}
\end{align}
where 
\begin{align}
\LLR_{*1}^{(2)}&=\log \frac{p(y_i| u_{1,i}^{(1)}, u_{2,i}^{(1)}, s_{i}^{(2)}=0)}{p(y_i|  u_{1,i}^{(1)}, u_{2,i}^{(1)}, s_{i}^{(2)}=1)}.
\end{align}
The conditional probability density function that describes $\LLR_{*1}^{(2)}$ is given by
\begin{align}
\label{eqn:LLR12}
&p(y_{i}|s_{i}^{(2)},u_{2,i}^{(1)},u_{1,i}^{(1)})=\sum_{s_i^{(3:L)},u_{1,i}^{(2:L)}} p(y_i|u_{1,i},u_{2,i}=u_{1,i} \oplus_{2^L} s_i) \nonumber \\
&=\sum_{s_i^{(3:L)},u_{1,i}^{(2:L)}} \frac{1}{\sqrt{2 \pi}} \exp{\left(-\frac{1}{2}(y_i-\varphi_1(u_{1,i})-\varphi_2(u_{1,i} \oplus_{2^L} s_i))^2\right)}.
\end{align}

The conditional probability density function that describes $\LLR_2^{(2)}$ is given by
\begin{align}
\label{eqn:LLR22}
&p(y_{i}|u_{1,i}^{(1)},u_{2,i}^{(1)},s_{i}^{(2)},u_{1,i}^{(2)})=\sum_{s_i^{(3:L)},u_{1,i}^{(3:L)}} p(y_i|u_{1,i},u_{2,i}=u_{1,i} \oplus_{2^L} s_i) \nonumber \\
&=\sum_{s_i^{(3:L)},u_{1,i}^{(3:L)}} \frac{1}{\sqrt{2 \pi}} \exp{\left(-\frac{1}{2}(y_i-\varphi_1(u_{1,i})-\varphi_2(u_{1,i} \oplus_{2^L} s_i))^2\right)}.
\end{align}

The conditional probability density function necessary for deriving $\LLR_{1}^{(\ell)}$ is
\begin{align}
&p(y_{i}|s_{i}^{(\ell)}, u_{1,i}^{(1:\ell-1)}, u_{2,i}^{(1:\ell-1)})=\sum_{s_i^{(\ell +1:L)},u_{1,i}^{(\ell :L)}} p(y_i|u_{1,i},u_{2,i}=u_{1,i} \oplus_{2^L} s_i) \nonumber \\
&=\sum_{s_i^{(\ell +1:L)},u_{1,i}^{(\ell:L)}} \frac{1}{\sqrt{2 \pi}} \exp{\left(-\frac{1}{2}(y_i-\varphi_1(u_{1,i})-\varphi_2(u_{1,i} \oplus_{2^L} s_i))^2\right)}.
\end{align}


The conditional probability density function necessary for deriving $\LLR_2^{(\ell)}$ is
\begin{align}
&p(y_{i}|s_{i}^{(\ell)}, u_{1,i}^{(1:\ell)}, u_{2,i}^{(1:\ell-1)})=\sum_{s_i^{(\ell +1:L)},u_{1,i}^{(\ell+1 :L)}} p(y_i|u_{1,i},u_{2,i}=u_{1,i} \oplus_{2^L} s_i) \nonumber \\
&=\sum_{s_i^{(\ell +1:L)},u_{1,i}^{(\ell+1:L)}} \frac{1}{\sqrt{2 \pi}} \exp{\left(-\frac{1}{2}(y_i-\varphi_1(u_{1,i})-\varphi_2(u_{1,i} \oplus_{2^L} s_i))^2\right)}.
\end{align}


\bibliographystyle{IEEEtran}
\bibliography{IEEEabrv,nit}

\end{document}